\documentclass[preprint,12pt]{elsarticle}
\usepackage{}
\usepackage{floatrow}

\newfloatcommand{figurebox}{figure}[\nocapbeside][\dimexpr(\textwidth-\columnsep)/2\relax]
\newfloatcommand{tablebox}{table}[\nocapbeside][\dimexpr(\textwidth-\columnsep)/2\relax]
\usepackage{booktabs}
\usepackage{threeparttable}
\usepackage{amscd}
\usepackage{amssymb}
\usepackage{amsmath}
\usepackage{amsfonts}
\usepackage[pagewise]{lineno}
\usepackage{float}
\usepackage{algorithm,algorithmic}

\usepackage{eqparbox}

\usepackage{float}  
\usepackage{subfigure}  
\usepackage{graphicx}
\usepackage{amsmath,amssymb,graphicx,epstopdf}
\usepackage{algorithm,algorithmic}
\usepackage{subfigure}
\usepackage{caption}

\usepackage{array}
 \usepackage{tabularx}
\usepackage{multirow}
 \usepackage{multicol}
 \usepackage{arydshln}

\newtheorem{theorem}{Theorem}
\newtheorem{definition}{Definition}

\def\max{\mbox{max}\,}

\usepackage{amssymb}

\journal{Nuclear Physics B}

\begin{document}

\begin{frontmatter}



\title{Weighted Truncated Nuclear Norm Regularization for Low-Rank Quaternion Matrix Completion}
\author[lab1]{Liqiao Yang}
\ead{liqiaoyoung@163.com}
\author[lab1]{Kit Ian Kou\corref{cor1}}
\ead{kikou@umac.mo}
\author[lab2]{Jifei Miao}
\ead{jifmiao@163.com}

\address[lab1]{Department of Mathematics, Faculty of
	Science and Technology, University of Macau, Macau 999078, China}

\cortext[cor1]{Corresponding author}

\begin{abstract}
In recent years, quaternion matrix completion (QMC) based on low-rank regularization has been gradually used in image de-noising and de-blurring. Unlike low-rank matrix completion (LRMC) which handles RGB images by recovering each color channel separately, the QMC models utilize the connection of three channels by processing them as a whole. Most of the existing quaternion-based methods formulate low-rank QMC (LRQMC) as a quaternion nuclear norm (a convex relaxation of the rank) minimization problem.
 The main limitation of these approaches is that the singular values being minimized simultaneously so that the low-rank property could not be approximated well and efficiently. To achieve a more accurate low-rank approximation, the matrix-based truncated nuclear norm has been proposed and also been proved to have the superiority. In this paper, we introduce a quaternion truncated nuclear norm (QTNN) for LRQMC and utilize the alternating direction method of multipliers (ADMM) to get the optimization. We further propose weights to the residual error quaternion matrix during the update process for accelerating the convergence of the QTNN method with admissible performance. The weighted method utilizes a concise gradient descent strategy which has a theoretical guarantee in optimization.  The effectiveness of our method is illustrated by experiments on real visual data sets.

\end{abstract}




\end{frontmatter}


\section{Introduction}
\label{sec:intro}
Matrix completion (MC) methods are designed for recovering an incomplete matrix by utilizing limited information. With the innovation of MC strategies, MC has been utilized in many fields e.g., image completion \cite{5454406,6389682,DBLP:conf/stoc/JainNS13}, image classification \cite{DBLP:conf/aaai/LiuLTXW15,DBLP:conf/nips/CabralTCB11} and so on. It is also had been proved that if the rank of the image is high, it frequently means that this is a fairly noisy image \cite{6138863,8066348}, so in the area of image processing, the low-rank restriction could be used to achieve image de-noising and repair. Mathematically, the LRMC problem can be formulated as follows
\begin{equation}\label{model1}
   \min\limits_{\mathbf{X}} \text{rank} (\mathbf{X}),  \qquad  \text{ s.t.}   \quad   P_\Omega(\mathbf{X}-\mathbf{M})=0,
  \end{equation}
  where $\mathbf{X},\ \mathbf{M}\in \mathbb{R}^{m\times n}$, $\Omega$ is the set of the coordinate position of observed elements in observed matrix $\mathbf{M}$, $P_\Omega$ is a linear operator:
$$P_\Omega(\mathbf{(X-M)}_{ij})=
\begin{cases}
1& \text{$(i,j) \in \Omega$ }\\
0               & \text{$ (i,j) \not\in \Omega$ }.
\end{cases}$$

Nonetheless, the rank function is discrete and nonconvex, so solving problem (\ref{model1}) is a challenging process, and minimizing the rank function directly is NP-hard \cite{1384521}. Basing on the nuclear norm (NN) is the tightest convex surrogate of the matrix rank, at one time the NN of matrix as the substitute for the rank function has been widely studied for optimizing LRMC problems \cite{recht2010guaranteed,koltchinskii2011nuclear}:
\begin{equation}\label{model2}
   \min\limits_{\mathbf{X}}  \parallel \mathbf{X}\parallel_*    \qquad   \text{s.t.}   \quad   P_\Omega(\mathbf{X}-\mathbf{M})=0,
\end{equation}
where $\parallel \mathbf{X}\parallel_* $ is used to approximate the rank of the matrix $\mathbf{X}$. $\parallel \mathbf{X}\parallel_*=\sum_{i=1}^{\min(m,n)}\sigma_{i}(\mathbf{X})$, and $\sigma_{i}(\mathbf{X})$ is the i\emph{th} largest singular value of $\mathbf{X}$.
Comparing with the rank function where every nonzero singular value makes an equal contribution, the smaller the singular value is, the less contribution could be made in the nuclear norm. Hence, in real applications, the NN may not estimate the rank function with great precision. Besides, it is hard to satisfy the theoretical requirements of the NN heuristic in practice \cite{DBLP:journals/cacm/CandesR12}.

To get more progress in handling the LRMC problems, some authors made general changes on the nuclear norm, such as Hu et al. \cite{6389682} developed a truncated nuclear norm (TNN) to get a better recovery results, their models can be written as:
\begin{equation}\label{model3}
   \min\limits_{\mathbf{X}}\parallel\mathbf{X}\parallel_r    \qquad   \text{s.t.}   \quad   P_\Omega(\mathbf{X}-\mathbf{M})=0,
\end{equation}
where $\parallel\mathbf{X}\parallel_r=\sum_{i=r+1}^{\min(m,n)}\sigma_{i}(\mathbf{X})$, it is the sum of $\min(m,n)-r$ minimum singular values of $\mathbf{X}$.

Instead of solving model (\ref{model2}) directly, Hu et al. only minimized the smallest $\min(m,n)-r$ singular values since the values of the largest $r$ nonzero singular values would not influence the process of determining the rank of the matrix. Therefore, they let these singular values free and focused on minimizing the sum of the smallest $\min(m,n)-r$ singular values. The main step in this strategy is to minimize the truncated nuclear norm of the target matrix, and the specific operation can be found in \ref{sec:2.1}.

Although the above TNN method had been shown better performance and faster convergence than strategies base on NN, when processing color images the most matrix completion methods only can deal with each channel separately and then combine the results together. It means that the LRMC methods could obtain remarkable recovery in the two-dimensional fashion, specifically, dealing with the grayscale images. So, when dealing with color images in order to avoid breaking the structure of the RGB channels and fully utilize their correlation among three channels, the quaternion-based representation for color images gradually entered into the field of study.

A quaternion is comprised of a real part and three imaginary components \cite{zhang1997quaternions}, and this structure is suitable for color image processing exactly. Using quaternion matrix, a color image can be represented by a pure quaternion matrix, and in this way, a color image can be treated as a whole instead of separating each RGB channels and dealing with them individually. Inspiring by the advantages of quaternion representation, many quaternion-based algorithms been designed for various fields of image processing. For example, in \cite{WANG2013255} a robust color image watermarking method been designed, in \cite{7024169} a modified vector sparse representation model for color image denoising algorithm been developed, which is based on \cite{4483511}, and an efficient color face recognition strategy been proposed in \cite{7468463}. In the field of color image inpainting, Chen et al. \cite{8844978} extended the traditional LRMC model in (\ref{model2}) to the quaternion-based model, they developed a low-rank quaternion matrix completion (LRQMC) model and use three nonconvex functions to improve the nuclear norm of the quaternion matrix. More precisely, in \cite{8844978}, all the nonnegative singular values of the quaternion matrix been adjusted to get better results, besides, it is necessary for this strategies to compute a large quaternion singular value decomposition (QSVD) in every step. The computations of large-size QSVD in each iteration is expensive, therefore the authors in \cite{9204671} based on quaternion double Frobenius norm (Q-DFN), quaternion double nuclear norm (Q-DNN), and quaternion Frobenius/nuclear norm (Q-FNN), designed a novel LRQMC method by recovering two smaller factor quaternion matrices to obtain the recovered results. In this LRQMC method, the process of large-scale QSVD has been avoided, so this strategy would reduce the time consumption.

While the above LRQMC methods took advantage of QTNN, all the nonnegative singular values been handled in the optimization process. Being inspired by the TNN in matrix cases, we extended it to quaternion domain named QTNN for more accurate results and time-saving. Further, we develop a weighted method for making effective use of the missing cases and speed up the convergence with admissible performance. In summary, the main contributions of this paper include:

\begin{itemize}
 \item We propose a new approach called the quaternion truncated nuclear norm (QTNN) for LRQMC. The rank of quaternion matrix would not be influenced by the largest $r$ singular values, so the proposed QTNN achieves a more accurate approximation for the rank function. In comparison with traditional TNN methods which regard color images as three matrices for LRMC, we regard the color image as a quaternion matrix and preserve the information well among three channels for LRQMC.

 \item  For speeding up the computation of the QTNN, we further propose double-weighted quaternion truncated nuclear norm (DWQTNN) strategy by adding weighted real diagonal matrices on the residual error quaternion matrix in augmented Lagrange function. When some missing rows have been recovered or roughly recovered, the completion would be much easier, so these weights aim to make the less missing rows of the missing quaternion matrix be recovered with higher accuracy than other rows. Moreover, we utilize the gradient descent method to get the recovery and prove the convergence in quaternion domain. This gradient search method is faster and more concise than the two-step approach we use in QTNN.

 \item In experiments, both of the parameters results and the outcomes of recovering missing color images demonstrate that our method is competitive.
 \end{itemize}

 The structure of the reminder paper is organized as follows. Section \ref{sec:1} presents the notations and preliminary knowledge about quaternion. Section \ref{sec:2} introduces the quaternion-based model and gives our method for completion. In Section \ref{Experimental results}, numerical simulations are designed to clarify the performance of the proposed algorithm. Finally, a conclusion will be given in Section \ref{Conclusion}.

\section{Notations and preliminaries}\label{sec:1}

In this section, we briefly explain some main notations and summarize some preliminary quaternion-based knowledge that we used in this paper.

\subsection{Notations}
Quaternion space as an extension of the real space $\mathbb{R}$ and the complex space $\mathbb{C}$ is denoted by $\mathbb{H}$. In this paper,
we denote scalars as lowercase letters (e.g., a), vectors as boldface letters (e.g., $\textbf{a}$), matrices as boldface uppercase letters (e.g., $\textbf{A}$). Both $\textbf{I}_{r\times r}$ and $\textbf{I}_{r}$ represent the $r\times r$ identity matrix. A dot above the variables (e.g., $\dot{a}$, $\dot{\textbf{a}}$, $\dot{\textbf{A}}$) are used to denote the variables (e.g., scalars, vectors, matrices) in quaternion space $\mathbb{H}$, and $\Re(*)$ is the real part of the quaternion $*$.  $\mathbf{(\cdot)}^T$, $\mathbf{(\cdot)}^*$, $\mathbf{(\cdot)}^H$ and $\mathbf{(\cdot)}^{-1}$ represent the transpose, conjugate, conjugate transpose and the inverse of $\mathbf{\cdot}$, respectively. $|\cdot|$, $\|\cdot\|_F$, $\|\cdot\|_*$ are the absolute value or modulus, Frobenius norm, and nuclear norm of $\cdot$, respectively. The inner product of $(\circ_1)$ and $(\circ_2)$ is defined as $\langle\circ_1\cdot\circ_2\rangle\triangleq \text{tr}(\circ_1^H\circ_2)$, and tr$(\cdot)$ is the trace function.

\subsection{Preliminary quaternion-based knowledge}
Several important definitions and properties of quaternion numbers and matrices are introduced in the following:
Let a quaternion $\dot{a}\in\mathbb{H}$. $\dot{a}$ is defined as $\dot{a}=a_0+a_1\emph{i}+a_2\emph{j}+a_3\emph{k}$, where $a_l\in\mathbb{R} (l=0,1,2,3)$, and \emph{i, j, k} are three imaginary number units. A real part $\mathfrak{R}(\dot{a}) \triangleq a_0$ and an imaginary part $\mathfrak{I}(\dot{a}) \triangleq a_1\emph{i}+a_2\emph{j}+a_3\emph{k}$ make up the quaternion $\dot{a}$. If $a_0 = 0$, $\dot{a}$ is called a pure quaternion. $\dot{a}^{*} = a_0-a_1\emph{i}-a_2\emph{j}-a_3\emph{k}$ is the conjugate of $\dot{a}$ and $|\dot{a}|= \sqrt{\dot{a}\dot{a}^{*}}=\sqrt{a_0^2+a_1^2+a_2^2+a_3^2}$ is the modulus of $\dot{a}$. Let two quaternions $\dot{a}$ and $\dot{b}\in\mathbb{H}$, $c\in\mathbb{R}$:

\begin{equation}\nonumber
   \dot{a}+\dot{b}=(a_0+b_0)+(a_1+b_1)\emph{i}+(a_2+b_2)\emph{j}+(a_3+b_3)\emph{k}
\end{equation}

\begin{equation}\nonumber
   c\dot{a}=(ca_0)+(ca_1)\emph{i}+(ca_2)\emph{j}+(ca_3)\emph{k}
\end{equation}

\begin{equation}\nonumber
 \begin{aligned}
   \dot{a}\dot{b}=& (a_0b_0-a_1b_1-a_2b_2-a_3b_3)\\&+(a_0b_1+a_1b_0+a_2b_3-a_3b_2)\emph{i}
   \\&+(a_0b_2-a_1b_3+a_2b_0+a_3b_1)\emph{j}\\&+(a_0b_3+a_1b_2-a_2b_1+a_3b_0)\emph{k}
 \end{aligned}
\end{equation}
It should be noted that the multiplication in quaternion is not commutative, $\dot{a}\dot{b} \neq \dot{b}\dot{a}$.

Let a quaternion matrix $\dot{\textbf{A}}=(\dot{a}_{ij})\in\mathbb{H}^{M \times N}$. $\dot{\textbf{A}}$ is defined as $\dot{\textbf{A}}=\textbf{A}_0+\textbf{A}_1\emph{i}+\textbf{A}_2\emph{j}+\textbf{A}_3\emph{k}$, where $\textbf{A}_l\in\mathbb{R}^{M \times N}(l=0,1,2,3)$, and if $\textbf{A}_0 = \textbf{0}$, $\dot{\textbf{A}}$ is called a pure quaternion matrix. $\parallel\dot{\textbf{A}}\parallel_F = \sqrt{\sum_{i=1}^{M}\sum_{j=1}^{N}|\dot{a}_{ij}|^2}=\sqrt{tr(\dot{\textbf{A}}^H\dot{\textbf{A}})}$. The expression using Cayley-Dickson notation \cite{DBLP:journals/sigpro/BihanM04} is $\dot{\textbf{A}}=\textbf{A}_p+\textbf{A}_q\emph{j}$, where $\textbf{A}_p$ and $\textbf{A}_q\in\mathbb{C}^{M \times N}$ are two complex matrices. Then the quaternion matrix $\dot{\textbf{A}}$ can be represented as an equivalent complex matrix $\textbf{A}_c\in\mathbb{C}^{2M \times 2N}$:
\begin{equation}
\textbf{A}_c={
\left( \begin{array}{cc}
\textbf{A}_p & \textbf{A}_q  \\
-\textbf{A}_q^* & \textbf{A}_p^*\\
\end{array}
\right )_{2M \times 2N}.}
\end{equation}

\begin{theorem}(\textbf{QSVD} \cite{zhang1997quaternions}):
\label{theorem1}
 For any quaternion matrix $\dot{\textbf{A}}\in\mathbb{H}^{M \times N}$ of rank
r, there are two unitary quaternion matrices $\dot{\textbf{U}}\in\mathbb{H}^{M \times M}$
and $\dot{\textbf{V}}\in\mathbb{H}^{N \times N}$ such that
\begin{equation}
\dot{\textbf{A}}={\dot{\textbf{U}}
\left( \begin{array}{cc}
\mathbf{\Sigma}_r & \mathbf{0}  \\
\mathbf{0} & \mathbf{0}\\
\end{array}
\right )\dot{\textbf{V}}^H,}
\end{equation}
where $\mathbf{\Sigma}_r=diag({\sigma_1,\cdots, \sigma_r})\in\mathbb{R}^{r\times r}$, and all singular values $\sigma_i (i=1,\cdots,r)$ are nonnegative.
\end{theorem}

Base on the above property and theorem, the QSVD can be obtained by computing the classical SVD of the complex matrix $\textbf{A}_c$, and more details can be found in \cite{9204671}. For detailed introduction of quaternion algebra, please refer to \cite{zhang1997quaternions, girard2007quaternions}. Following the Theorem \ref{theorem1}, we can get the definitions of quaternion rank and QNN:

\begin{definition}(\textbf{Quaternion rank}\cite{8844978}): The number of nonzero singular values is the rank of the quaternion matrix $\dot{\textbf{A}}\in\mathbb{H}^{M \times N}$.
\end{definition}

\begin{definition}(\textbf{QNN}\cite{8844978}): Given $\dot{\textbf{A}}\in\mathbb{H}^{M \times N}$, the sum of all nonzero singular values is the nuclear norm of the quaternion matrix, i.e., $\parallel\dot{\textbf{A}}\parallel_*=\sum_{i=1}^{min(M,N)}\sigma_{i}(\dot{\textbf{A}})$.
\end{definition}

It has been proved in \cite{zhang1997quaternions} that QSVD has similar forms and properties of the SVD in real domain, like the nonnegativity that above-mentioned and the decreasing order property of singular values. Hence, we define the quaternion truncated nuclear norm which derived from the truncated nuclear norm in real domain \cite{6389682} as following:

\begin{definition}(\textbf{QTNN }):
 \label{definition3}
 Given $\dot{\textbf{A}}\in\mathbb{H}^{M \times N}$, the sum of $min(M,N)-r$ minimum singular values is the quaternion truncated nuclear norm of the quaternion matrix, i.e., $\parallel\dot{\textbf{A}}\parallel_r=\sum_{i=r+1}^{min(M,N)}\sigma_{i}(\dot{\textbf{A}})$.
\end{definition}

\section{Low-rank completion model and algorithm }\label{sec:2}
In this section, we first introduce the specific optimization process of LRMC by using TNNR, then we propose our quaternion-based low-rank completion model and the optimization process by adding weights.

\subsection{Truncated Nuclear Norm Regularization Method (TNNR)}\label{sec:2.1}
As we briefly referred in the the model (\ref{model2}), the TNNR model is given by minimizing the sum of several smallest singular values. However, this truncated norm $\parallel\mathbf{X}\parallel_r$ is nonconvex, so the authors utilized the Von Neumann's trace inequality \cite{mirsky1975trace} to handle the nonconvexity. In this way, the TNNR model is formulated as follows:
\begin{equation}\label{model4}
\begin{aligned}
&\min\limits_{\mathbf{X}}\parallel\mathbf{X}\parallel_*- \min\limits_{\mathbf{C}\mathbf{C}^T=\mathbf{I},\mathbf{D}\mathbf{D}^T=\mathbf{I}} tr(\mathbf{C}\mathbf{X}\mathbf{D}^T)   \\& \text{s.t.}   \quad   P_\Omega(\mathbf{X}-\mathbf{M})=0,
\end{aligned}
\end{equation}
where $\mathbf{C}\in\mathbb{R}^{r\times m}$, $\mathbf{D}\in\mathbb{R}^{r\times n}$, and r is the truncated number.

There are two main steps at the process of solving model (\ref{model4}). Setting $\mathbf{X}_k$ is the \emph{k}th iteration of $\mathbf{X}$, $\mathbf{C}_k$ and $\mathbf{D}_k$ are obtained by computing the SVD of $\mathbf{X}_k$ at the first step. Then in step 2, solving $\mathbf{X}_{k+1}$ by the following equation:
\begin{equation}
arg\min\limits_{\mathbf{X}}\parallel\mathbf{X}\parallel_*-tr(\mathbf{C}_k\mathbf{X}\mathbf{D}_k^T)
\end{equation}
The alternating direction method of multipliers (ADMM) method is efficient to solve the subproblem for the convergence guarantee
\cite{boyd2011distributed}, and the main procedure is summarized in Table \ref{table1}.

\begin{table}[htbp]
\caption{The two-step TNNR algorithm}
\hrule
\label{table1}
\begin{algorithmic}[1]
\REQUIRE   the incomplete matrix data $\mathbf{M}\in\mathbb{R}^{m\times n}$, the position set of observed elements $\Omega$, and the tolerance $\varepsilon_0$.
\STATE \textbf{Initial} $\mathbf{X}_1=P_\Omega(\mathbf{M})$.
\STATE \textbf{Repeat}
\STATE \quad \textbf{Step 1.} Given $\mathbf{X}_{k}$\\
  \qquad  \qquad \qquad  $[\mathbf{U}_{k},\mathbf{\Sigma}_{k},\mathbf{V}_{k}]=SVD(\mathbf{X}_{k})$
\STATE  \quad where $\mathbf{U}_{k}=(\mathbf{u}_{1},\cdots\mathbf{u}_{m})\in\mathbb{R}^{m\times m}$,\\
 \qquad \quad\quad $\mathbf{V}_{k}=(\mathbf{v}_{1},\cdots\mathbf{v}_{n})\in\mathbb{R}^{n\times n}$.
\STATE   \quad Computing $\mathbf{C}_k=(\mathbf{u}_{1},\cdots\mathbf{u}_{r})^T\in\mathbb{R}^{r\times m}$ and \\ \qquad\qquad\quad\quad$\mathbf{D}_k=(\mathbf{v}_{1},\cdots\mathbf{v}_{r})^T\in\mathbb{R}^{r\times n}$.
\STATE \quad \textbf{Step 2.} Solving  \\
 \qquad \quad\quad $\mathbf{X}_{k+1}=arg\min\parallel\mathbf{X}\parallel_*-tr(\mathbf{C}_k\mathbf{X}\mathbf{D}_k^T)$
\STATE \textbf{Until convergence} $\|\mathbf{X}_{k+1}-\mathbf{X}_{k}\|_F \leq \varepsilon_0$
\ENSURE  the recovered matrix.
\end{algorithmic}
\hrule
\end{table}
\subsection{The proposed Quaternion-based Truncated Nuclear Norm Regularization Method (QTNNR)}

For the LRMC model, it reveals an obvious drawback that the matrix-based method processes each channel separately, and ignores the inter-relationship among the three channels. This optimization is likely to produce distortions in the reconstruction results for color images, while if regarding a color image as a quaternion matrix and optimizing this inpainting problem in quaternion domain, the relationship among each channel  and the inherent color structures would be preserved. Hence, recently more studies formulate the inpainting problem as a low-rank quaternion completion problem, and our LRQMC model can be defined as follows.

Based on Definition \ref{definition3}, we develop the LRQMC model by quaternion truncated nuclear norm as follows:
\begin{equation}\label{model5}
   \min\limits_{\dot{\mathbf{X}}}\parallel\dot{\mathbf{X}}\parallel_r    \qquad   \text{s.t.}   \quad   P_\Omega(\dot{\mathbf{X}}-\dot{\mathbf{M}})=0,
\end{equation}
where $\|\dot{\mathbf{X}}\|_r=\sum_{i=r+1}^{\min(m,n)}\sigma_{i}(\dot{\mathbf{X}})$, it is the sum of the $\min(m,n)-r$ minimum singular values of $\dot{\mathbf{X}}$. The $P_\Omega$ is same like in (\ref{model1}).

For solving (\ref{model5}), we have the following theorem:
\begin{theorem}
\label{theorem3}
For any quaternion matrix $\dot{\textbf{X}}\in\mathbb{H}^{M \times N}$, and any matrices $\dot{\textbf{A}}\in\mathbb{H}^{r \times M}$  and $\dot{\textbf{B}}\in\mathbb{H}^{r \times N}$ that are satisfied with $\dot{\textbf{A}}\dot{\textbf{A}}^{H}=\textbf{I}_{r\times r}$, $\dot{\textbf{B}}\dot{\textbf{B}}^{H}=\textbf{I}_{r\times r}$. r is any nonnegative integer $(r\leq min(M,N))$, we have
\begin{equation}
\mid tr(\dot{\textbf{A}}\dot{\mathbf{X}}\dot{\mathbf{B}}^{H})\mid\leq\sum_{i=1}^r\sigma_{i}(\dot{\mathbf{X}}).
\end{equation}
Besides, we have
\begin{equation}
\begin{aligned}
\max |tr(\dot{\mathbf{A}}\dot{\mathbf{X}}\dot{\mathbf{B}}^H)|=\sum_{i=1}^r\sigma_{i}(\dot{\mathbf{X}}).
\end{aligned}
\end{equation}
\end{theorem}
The proof of Theorem \ref{theorem3} can be found in the Appendix \ref{A1}.

Then the problem (\ref{model5}) can be rewritten as:
\begin{equation}\label{model6}
\begin{aligned}
&\min\limits_{\dot{\mathbf{X}}}\parallel\dot{\mathbf{X}}\parallel_*- \mathop{\max}\limits_{\dot{\mathbf{C}}\dot{\mathbf{C}}^H=\mathbf{I},\dot{\mathbf{D}}\dot{\mathbf{D}}^H=\mathbf{I}} |tr(\dot{\mathbf{C}}\dot{\mathbf{X}}\dot{\mathbf{D}}^H)|  \qquad   \\& \text{s.t.}   \quad   P_\Omega(\dot{\mathbf{X}}-\dot{\mathbf{M}})|=0,
\end{aligned}
\end{equation}
where $\dot{\mathbf{C}}=(\dot{\mathbf{u}}_{1},\cdots \dot{\mathbf{u}}_{r})^H$ and $\dot{\mathbf{D}}=(\dot{\mathbf{v}}_{1},\cdots \dot{\mathbf{v}}_{r})^H$, and $\{\dot{\mathbf{u}}_{1},\cdots\dot{\mathbf{u}}_{r}\}, \{\dot{\mathbf{v}}_{1},\cdots\dot{\mathbf{v}}_{r}\}$ are the first r columns of the quaternion matrices $\dot{\mathbf{U}}$, $\dot{\mathbf{V}}$ separately. $\dot{\mathbf{U}}$, $\dot{\mathbf{V}}$ are left and right unitary quaternion matrices from operating QSVD  on $\dot{\mathbf{X}}$.

Inspired by the TNNR optimized method, at the $l$\emph{th} iteration, in \textbf{Step 1} we obtained $\dot{\mathbf{C}}_l$ and $\dot{\mathbf{D}}_l$ by operating QSVD on $\dot{\mathbf{X}}_l$, then in \textbf{Step 2} we fix $\dot{\mathbf{C}}_l$, $\dot{\mathbf{D}}_l$ and obtain $\dot{\mathbf{X}}_{l+1}$ by solving the following subproblem:
\begin{equation}\label{model6b}
\begin{aligned}
&\min\limits_{\dot{\mathbf{X}}}\parallel\dot{\mathbf{X}}\parallel_*- |tr(\dot{\mathbf{C}}_l\dot{\mathbf{X}}\dot{\mathbf{D}}_l^H)|  \qquad   \\& \text{s.t.}   \quad   P_\Omega(\dot{\mathbf{X}}-\dot{\mathbf{M}})=0,
\end{aligned}
\end{equation}
we use ADMM frame to optimize the subproblem (\ref{model6b}).

Firstly, adding an intermediate variable $\dot{\mathbf{H}}$, (\ref{model6b}) can be rewritten as:
\begin{equation}\label{model7}
\begin{aligned}
&\min\limits_{\dot{\mathbf{X}},\dot{\mathbf{H}}}\parallel\dot{\mathbf{X}}\parallel_*- |tr(\dot{\mathbf{C}}_l\dot{\mathbf{H}}\dot{\mathbf{D}}_l^H)|  \qquad   \\& \text{s.t.}  \quad \dot{\mathbf{X}}=\dot{\mathbf{H}} \quad  P_\Omega(\dot{\mathbf{H}}-\dot{\mathbf{M}})=0.
\end{aligned}
\end{equation}
Because the quaternion multiplication is not commutative and in analogy with such cases in \cite{9204671}, the augmented Lagrange function of (\ref{model7}) is:
\begin{equation}\label{model8}
\begin{split}
L(\dot{\mathbf{X}},\dot{\mathbf{H}},\dot{\mathbf{Y}}, \beta_k)=&\parallel\dot{\mathbf{X}}\parallel_*- |tr(\dot{\mathbf{C}}_l\dot{\mathbf{H}}\dot{\mathbf{D}}_l^H)|\\&
+\frac{\beta_k}{2}\parallel\dot{\mathbf{X}}-\dot{\mathbf{H}}\parallel_F^2\\&
+\mathfrak{R}(tr(\dot{\mathbf{Y}}^H(\dot{\mathbf{X}}-\dot{\mathbf{H}}))),
\end{split}
\end{equation}
where $\beta_k>0$ is the penalty parameter, and $\dot{\mathbf{Y}}$ is the Lagrange multiplier. Set $\dot{\mathbf{X}}_1=P_\Omega(\dot{\mathbf{M}})$, $\dot{\mathbf{H}}_1 = \dot{\mathbf{X}}_1$, and $\dot{\mathbf{Y}}_1 = \dot{\mathbf{X}}_1$
as the initialization in \textbf{Step 2}. The optimization of (\ref{model8}) consists of the following four steps:

\textbf{Step 2.1:} Updating $\dot{\mathbf{X}}_{k+1}$. Keeping $\dot{\mathbf{H}}_k$ and $\dot{\mathbf{Y}}_k$ fixed, and minimizing $L(\dot{\mathbf{X}},\dot{\mathbf{H}}_k,\dot{\mathbf{Y}}_k, \beta)$ as follows:
\begin{equation}\label{model9}
\begin{split}
\dot{\mathbf{X}}_{k+1}&=arg\min_{\dot{\mathbf{X}}}L(\dot{\mathbf{X}},\dot{\mathbf{H}}_k,\dot{\mathbf{Y}}_k, \beta_k)\\&=arg\min_{\dot{\mathbf{X}}}\parallel\dot{\mathbf{X}}\parallel_*- |tr(\dot{\mathbf{C}}_l\dot{\mathbf{H}}\dot{\mathbf{D}}^H_l)|
\\& \quad +\frac{\beta_k}{2}\parallel\dot{\mathbf{X}}-\dot{\mathbf{H}}_k\parallel_F^2
+\mathfrak{R}(tr(\dot{\mathbf{Y}}^H_k(\dot{\mathbf{X}}-\dot{\mathbf{H}}_k))),
\end{split}
\end{equation}
Ignoring the constant terms, (\ref{model9}) can be rewritten as:
\begin{equation}\label{model10}
\begin{split}
\dot{\mathbf{X}}_{k+1}=arg\min_{\dot{\mathbf{X}}}\parallel\dot{\mathbf{X}}\parallel_*+
\frac{\beta}{2}\parallel\dot{\mathbf{X}}-(\dot{\mathbf{H}}_k-\frac{1}{\beta_k}\dot{\mathbf{Y}}_k)\parallel_F^2.
\end{split}
\end{equation}
Based on quaternion singular value thresholding (QSVT) \cite{8844978} which enjoys similar forms of the singular value thresholding (SVT) \cite{cai2010singular} in real matrix domain. In this way, we can solve (\ref{model10}) efficiently and obtain the closed solution by:
\begin{equation}\label{model11}
\dot{\mathbf{X}}_{k+1}=\mathfrak{D}_{\frac{1}{\beta_k}}(\dot{\mathbf{H}}_k-\frac{1}{\beta_k}\dot{\mathbf{Y}}_k).
\end{equation}

\textbf{Step 2.2:} Updating $\dot{\mathbf{H}}_{k+1}$. Keeping $\dot{\mathbf{X}}_{k+1}$ and $\dot{\mathbf{Y}}_k$ fixed, and minimizing $L(\dot{\mathbf{X}}_{k+1},\dot{\mathbf{H}},\dot{\mathbf{Y}}_k, \beta_k)$ as follows:
\begin{equation}\label{model12}
\begin{aligned}
\dot{\mathbf{H}}_{k+1}&=arg\min_{\dot{\mathbf{H}}}L(\dot{\mathbf{X}}_{k+1},\dot{\mathbf{H}},\dot{\mathbf{Y}}_k, \beta_k)\\&=arg\min_{\dot{\mathbf{H}}}- |tr(\dot{\mathbf{C}}_l\dot{\mathbf{H}}\dot{\mathbf{D}}_l^H)|
+\frac{\beta_k}{2}\parallel\dot{\mathbf{X}}_{k+1}-\dot{\mathbf{H}}\parallel_F^2\\& \quad
+\mathfrak{R}(tr(\dot{\mathbf{Y}}^H_k(\dot{\mathbf{X}}_{k+1}-\dot{\mathbf{H}}))).
\end{aligned}
\end{equation}
 Discarding the constant terms in (\ref{model12}), it can be reformulated as:
 \begin{equation}\label{model12b}
\dot{\mathbf{H}}_{k+1}=arg\min_{\dot{\mathbf{H}}}\frac{\beta_k}{2}\|\dot{\mathbf{H}}-(\dot{\mathbf{X}}_{k+1}+
\frac{1}{\beta_k}(\dot{\mathbf{C}}_l^H\dot{\mathbf{D}}_l+\dot{\mathbf{Y}}_k))\|^2_F
\end{equation}
, and the closed  form solution of $\dot{\mathbf{H}}_{k+1}$ can be obtained by:
 \begin{equation}\label{model13}
\dot{\mathbf{H}}_{k+1}=\dot{\mathbf{X}}_{k+1}+\frac{1}{\beta_k}(\dot{\mathbf{C}}_l^H\dot{\mathbf{D}}_l+\dot{\mathbf{Y}}_k).
\end{equation}
Then, we let the values of all observed elements be constant in each iteration and obtain:
 \begin{equation}\label{model14}
\dot{\mathbf{H}}_{k+1}=P_{\Omega^C}(\dot{\mathbf{H}}_{k+1})+P_\Omega(\dot{\mathbf{M}}).
\end{equation}

\textbf{Step 2.3:} Updating $\dot{\mathbf{Y}}_{k+1}$. Keeping $\dot{\mathbf{X}}_{k+1}$ and $\dot{\mathbf{H}}_{k+1}$ fixed, and $\dot{\mathbf{Y}}_{k+1}$ can be obtained directly by:
\begin{equation}\label{model15}
\dot{\mathbf{Y}}_{k+1}=\dot{\mathbf{Y}}_{k}+\beta_k(\dot{\mathbf{X}}_{k+1}-\dot{\mathbf{H}}_{k+1}).
\end{equation}

\textbf{Step 2.4:} Updating the penalty parameter $\beta_{k+1}$ by:
\begin{equation}
\beta_{k+1}=\rho\beta_{k}.
\end{equation}

Although both the accurate and efficiency are getting progress by the above QTNN method, we can directly find that the process of recovering all the missing elements of the incomplete quaternion matrix is happening at the same time. In matrix cases, it has been observed that if some rows  with more observed are covered with higher accuracy, the completion task would be more easier \cite{liu2015truncated} and it would be time-saving. So in next subsection we develop two modified optimizations for improving \textbf{Step 2} with weighted matrices which are based on the gradient search, named weighted QTNN (WQTNN) and double weighted QTNN (DWQTNN), separately.
\subsection{ The proposed WQTNN and DWQTNN methods}
According to Theorem \ref{theorem3}, when the truncated number $r$ equals to  $ min(M,N)$, we have:
\begin{equation}
\mid tr(\dot{\textbf{A}}\dot{\mathbf{X}}\dot{\mathbf{B}}^{H})\mid\leq\sum_{i=1}^{\min(M,N)}\sigma_{i}(\dot{\mathbf{X}}),
\end{equation}
where $\dot{\textbf{A}}\dot{\textbf{A}}^H=\dot{\textbf{B}}\dot{\textbf{B}}^H=\textbf{I}_{\min(M,N)}$, and it means that
\begin{equation}
\max\mid tr(\dot{\textbf{A}}\dot{\mathbf{X}}\dot{\mathbf{B}}^{H})\mid=\|\dot{\mathbf{X}}\|_*,
\end{equation}
so the problem (\ref{model6}) can be reformulated as follows:
\begin{equation}\label{model16}
\begin{aligned}
\min\limits_{\dot{\mathbf{X}}}&\mathop{\max}\limits_{\dot{\textbf{A}}\dot{\textbf{A}}^H=\dot{\textbf{B}}\dot{\textbf{B}}^H=\textbf{I}_{\min(M,N)}}
|tr(\dot{\mathbf{A}}\dot{\mathbf{X}}\dot{\mathbf{B}}^H)|\\&- \mathop{\max}\limits_{\dot{\textbf{C}}\dot{\textbf{C}}^H=\dot{\textbf{D}}\dot{\textbf{D}}^H=\textbf{I}_{r}}
|tr(\dot{\mathbf{C}}\dot{\mathbf{X}}\dot{\mathbf{D}}^H)| \qquad   \\& \text{s.t.} \quad  P_\Omega(\dot{\mathbf{X}}-\dot{\mathbf{M}})=0.
\end{aligned}
\end{equation}
To simplify the derivation, we assume $\min(M,N)=M$ in the following. In \textbf{Step 1}, we can obtain quaternion matrices $\dot{\textbf{A}}_l,\dot{\textbf{B}}_l,\dot{\textbf{C}}_l,\dot{\textbf{D}}_l$ directly  at the $l$\emph{th} iteration, by the following equations:
\begin{equation}\label{model17}
\dot{\mathbf{A}}_l=\dot{\mathbf{U}}^H,
\end{equation}
\begin{equation}\label{model18}
\dot{\mathbf{B}}_l=(\dot{\mathbf{v}}_{1},\cdots\dot{\mathbf{v}}_{M})^H,
\end{equation}
\begin{equation}\label{model19}
\dot{\mathbf{C}}_l=(\dot{\mathbf{u}}_{1},\cdots\dot{\mathbf{u}}_{r})^H,
\end{equation}
\begin{equation}\label{model20}
\dot{\mathbf{D}}_l=(\dot{\mathbf{v}}_{1},\cdots\dot{\mathbf{v}}_{r})^H,
\end{equation}
where $\dot{\mathbf{U}}$ and $\dot{\mathbf{V}}$ are the left and right unitary quaternion matrices respectively that are obtained from QSVD of $\dot{\mathbf{X}}_l$, and r is the number of truncated singular values. Besides, in the following iteration, the matrices $\dot{\textbf{A}}_l,\dot{\textbf{B}}_l,\dot{\textbf{C}}_l,\dot{\textbf{D}}_l$ are fixed. In this way, the updating of $\dot{\mathbf{X}}_{l+1}$ can be formulated as:
\begin{equation}\label{model21}
\begin{aligned}
&\min\limits_{\dot{\mathbf{X}}}
|tr(\dot{\mathbf{A}}_l\dot{\mathbf{X}}\dot{\mathbf{B}}_l^H)|-
|tr(\dot{\mathbf{C}}_l\dot{\mathbf{X}}\dot{\mathbf{D}}_l^H)|   \qquad   \\& \text{s.t.}  \quad P_\Omega(\dot{\mathbf{X}}-\dot{\mathbf{M}})=0.
\end{aligned}
\end{equation}
The ADMM frame is adopted to solve (\ref{model21}) by adding an intermediate variable $\dot{\mathbf{H}}$ to relax the constraint, and (\ref{model21}) can be reformulated as:
\begin{equation}\label{model22}
\begin{aligned}
&\min\limits_{\dot{\mathbf{X}}}
|tr(\dot{\mathbf{A}}_l\dot{\mathbf{X}}\dot{\mathbf{B}}_l^H)|-
|tr(\dot{\mathbf{C}}_l\dot{\mathbf{H}}\dot{\mathbf{D}}_l^H)|   \qquad   \\& \text{s.t.}  \quad \dot{\mathbf{X}}=\dot{\mathbf{H}} \quad P_\Omega(\dot{\mathbf{H}}-\dot{\mathbf{M}})=0.
\end{aligned}
\end{equation}
Then the augmented lagrange function of (\ref{model22}) is:
\begin{equation}\label{model23}
\begin{split}
L(\dot{\mathbf{X}},\dot{\mathbf{H}},\dot{\mathbf{Y}}, \beta)=&|tr(\dot{\mathbf{A}}_l\dot{\mathbf{X}}\dot{\mathbf{B}}_l^H)|-
|tr(\dot{\mathbf{C}}_l\dot{\mathbf{H}}\dot{\mathbf{D}}_l^H)| \\&+\frac{\beta}{2}\parallel\dot{\mathbf{X}}-\dot{\mathbf{H}}\parallel_F^2+
\mathfrak{R}(tr(\dot{\mathbf{Y}}^H(\dot{\mathbf{X}}-\dot{\mathbf{H}}))),
\end{split}
\end{equation}
where $\beta_k>0$ is the penalty parameter, and $\dot{\mathbf{Y}}$ is the Lagrange multiplier.

As \cite{liu2015truncated} pointed out, the matrix completion problem would be much easier after some rows with missing entries have been recovered or roughly recovered. Hence, to further accelerate the convergence speed of QTNN, we add the weighted real matrix $\mathbf{W}$ to the residual error of $\dot{\mathbf{X}}$ and $\dot{\mathbf{H}}$. As a result, the augmented Lagrangian function (\ref{model23}) becomes:
\begin{equation}\label{model24}
\begin{split}
L(\dot{\mathbf{X}},\dot{\mathbf{H}},\dot{\mathbf{Y}}, \beta_k)=&\mathfrak{R}(tr(\dot{\mathbf{A}}_l\dot{\mathbf{X}}\dot{\mathbf{B}}_l^H))-
\mathfrak{R}(tr(\dot{\mathbf{C}}_l\dot{\mathbf{H}}\dot{\mathbf{D}}_l^H)) \\&+\frac{\beta_k}{2}\parallel\mathbf{W}(\dot{\mathbf{X}}-\dot{\mathbf{H}})\parallel_F^2
\\&+\mathfrak{R}(tr(\dot{\mathbf{Y}}^H(\mathbf{W}(\dot{\mathbf{X}}-\dot{\mathbf{H}}))),
\end{split}
\end{equation}
where $\mathbf{W}$ is located at the left side of $\dot{\mathbf{X}}-\dot{\mathbf{H}}$ to control the recovered accuracy and priority of $\dot{\mathbf{X}}$'s rows, and $\mathbf{W}=diag(w_1\cdots w_M)$, $w_i>0$ $(i=1\cdots M)$. Utilizing the number of observed entries ($n_i, i=1\cdots M$) in different rows, e.g., if $n_i\leq n_j$, we set the corresponding $w_i\leq w_j\leq1$ then we can get the weighted matrix. Suppose $\dot{\mathbf{\mathbb{X}}}_k$ is the k\emph{th} iteration in \textbf{Step 2}. Set $\dot{\mathbf{\mathbb{X}}}_1=\dot{\mathbf{X}}_l$, $\dot{\mathbf{H}}_1 = \dot{\mathbf{\mathbb{X}}}_1$, and $\dot{\mathbf{Y}}_1 = \dot{\mathbf{\mathbb{X}}}_1$
as the initialization. The optimization of subproblem (\ref{model24}) consists of the following four steps:

\textbf{Step 2.1:} Updating $\dot{\mathbf{H}}_{k+1}$. Keeping $\dot{\mathbf{\mathbb{X}}}_{k}$ and $\dot{\mathbf{Y}}_k$ fixed,and minimizing $L(\dot{\mathbf{X}}_{k},\dot{\mathbf{H}},\dot{\mathbf{Y}}_k, \beta_k)$ as follows:
\begin{equation}\label{model25}
\begin{aligned}
\dot{\mathbf{H}}_{k+1}&=arg\min_{\dot{\mathbf{H}}}L(\dot{\mathbf{\mathbb{X}}}_{k},\dot{\mathbf{H}},\dot{\mathbf{Y}}_k, \beta_k)\\&=arg\min_{\dot{\mathbf{H}}}|tr(\dot{\mathbf{C}}_l\dot{\mathbf{H}}\dot{\mathbf{D}}_l^H)|
+\frac{\beta_k}{2}\parallel \mathbf{W}(\dot{\mathbf{\mathbb{X}}}_{k}-\dot{\mathbf{H}}_k)\parallel_F^2\\& \quad
+\mathfrak{R}(tr(\dot{\mathbf{Y}}^H_k(\mathbf{W}(\dot{\mathbf{\mathbb{X}}}_{k}-\dot{\mathbf{H}})))).
\end{aligned}
\end{equation}
Similar to the above operation in (\ref{model12b}), the closed solution of $\dot{\mathbf{H}}_{k+1}$ can be obtained:
 \begin{equation}\label{model26}
\dot{\mathbf{H}}_{k+1}=\dot{\mathbf{\mathbb{X}}}_{k}+\frac{1}{\beta_k}(\mathbf{W}^{-2}\dot{\mathbf{C}}_l^H\dot{\mathbf{D}}_l+\mathbf{W}^{-1}\dot{\mathbf{Y}}_k),
\end{equation}
where $\mathbf{W}^{-2}=diag(w_1^{-2}\cdots w_M^{-2})$ and $\mathbf{W}^{-1}=diag(w_1^{-1}\cdots w_M^{-1})$. Because $P_\Omega(\dot{\mathbf{H}}-\dot{\mathbf{M}})=0$, we can obtain:
 \begin{equation}\label{model27}
\dot{\mathbf{H}}_{k+1}=P_{\Omega^C}(\dot{\mathbf{H}}_{k+1})+P_\Omega(\dot{\mathbf{M}}).
\end{equation}

\textbf{Step 2.2:} Updating $\dot{\mathbf{\mathbb{X}}}_{k+1}$. Keeping $\dot{\mathbf{H}}_{k+1}$ and $\dot{\mathbf{Y}}_k$ fixed, and minimizing $L(\dot{\mathbf{\mathbb{X}}},\dot{\mathbf{H}}_{k+1},\dot{\mathbf{Y}}_k, \beta_k)$ as follows:
\begin{equation}\label{model28}
\begin{split}
\dot{\mathbf{\mathbb{X}}}_{k+1}&=arg\min_{\dot{\mathbf{X}}}L(\dot{\mathbf{\mathbb{X}}},\dot{\mathbf{H}}_{k+1},\dot{\mathbf{Y}}_k, \beta)\\&=arg\min_{\dot{\mathbf{\mathbb{X}}}} |tr(\dot{\mathbf{A}}_l\dot{\mathbf{\mathbb{X}}}\dot{\mathbf{B}}^H_l)|
\\& \quad +\frac{\beta}{2}\parallel\mathbf{W}(\dot{\mathbf{\mathbb{X}}}-\dot{\mathbf{H}}_{k+1})+\frac{1}{\beta}\dot{\mathbf{Y}}_k\parallel_F^2.
\end{split}
\end{equation}
Discarding the constant terms, we can obtain the following equation:
\begin{equation}\label{model29}
\begin{split}
&\dot{\mathbf{\mathbb{X}}}_{k+1}\\&=arg\min_{\dot{\mathbf{\mathbb{X}}}}
\frac{\beta_k}{2}\parallel\mathbf{W}(\dot{\mathbf{\mathbb{X}}}-\dot{\mathbf{H}}_{k+1})+\frac{1}{\beta_k}\dot{\mathbf{Y}}_k +(\mathbf{W}^{-1}\dot{\mathbf{A}}_l^H\dot{\mathbf{B}}_l) \parallel_F^2 \\& =arg\min_{\dot{\mathbf{\mathbb{X}}}}
\frac{\beta_k}{2}\parallel\dot{\mathbf{\mathbb{X}}}-\dot{\mathbf{H}}_{k+1}+\frac{1}{\beta_k}(\mathbf{W}^{-2}\dot{\mathbf{A}}_l^H\dot{\mathbf{B}}_l+\mathbf{W}^{-1}\dot{\mathbf{Y}}_k) \parallel_F^2,
\end{split}
\end{equation}
which has a closed solution as:
\begin{equation}\label{model30}
\dot{\mathbf{\mathbb{X}}}_{k+1}=
\dot{\mathbf{H}}_{k+1}-\frac{1}{\beta_k}(\mathbf{W}^{-2}\dot{\mathbf{A}}_l^H\dot{\mathbf{B}}_l+\mathbf{W}^{-1}\dot{\mathbf{Y}}_k).
\end{equation}

\textbf{Step 2.3:} Updating $\dot{\mathbf{Y}}_{k+1}$. Keep $\dot{\mathbf{\mathbb{X}}}_{k+1}$ and $\dot{\mathbf{H}}_{k+1}$ fixed, and $\dot{\mathbf{Y}}_{k+1}$ can be obtained directly by:
\begin{equation}\label{model31}
\dot{\mathbf{Y}}_{k+1}=\dot{\mathbf{Y}}_{k}+\beta_k\mathbf{W}(\dot{\mathbf{\mathbb{X}}}_{k+1}-\dot{\mathbf{H}}_{k+1}),
\end{equation}

\textbf{Step 2.4:} Updating the penalty parameter $\beta_{k+1}$ by:
\begin{equation}
\beta_{k+1}=\rho\beta_{k}.
\end{equation}

In fact, we can find that the above process also need many iterations for computing $\dot{\mathbf{Y}}_{k+1}$, $\dot{\mathbf{\mathbb{X}}}_{k+1}$ and $\dot{\mathbf{H}}_{k+1}$ in each step, and based on the intrinsic structural correlations between (\ref{model26}) and (\ref{model30}), we can substitute $\dot{\mathbf{H}}_{k+1}$ into $\dot{\mathbf{\mathbb{X}}}_{k+1}$, then we get:
\begin{equation}\label{model32}
\dot{\mathbf{\mathbb{X}}}_{k+1}=
\dot{\mathbf{\mathbb{X}}}_k-\frac{1}{\beta_k}(\mathbf{W}^{-2}\dot{\mathbf{A}}_l^H\dot{\mathbf{B}}_l-\mathbf{W}^{-2}\dot{\mathbf{C}}_l^H\dot{\mathbf{D}}_l).
\end{equation}
For (\ref{model32}), we have the following theorem to obtain a one-step gradient descent method and guarantee the convergence performance.

\begin{theorem}
\label{theorem4}
If $0 < \beta_t < \beta_{t+1}$, for $t=1, 2,\cdots $, $\lim_{k\rightarrow\infty}\frac{1}{\beta_k}=0$, and when the step size $\beta_k$ of (\ref{model31}) be changed to a smaller positive number  $\gamma_k$ $(0<\gamma_k<\beta_k)$, besides, $\gamma_k$ need to satisfy: $\sum_{k=1}^{+\infty}\frac{\gamma_k}{\beta_k}=c$ (c is a positive constant). Based on the above optimizations (\ref{model27})-(\ref{model31}) and $\dot{\mathbf{X}}_l=\dot{\mathbf{\mathbb{X}}}_1$ , the consequence ${\dot{\mathbf{\mathbb{X}}}_k}$ will converge as the following form:
\begin{equation}
\dot{\mathbf{X}}_*=
\dot{\mathbf{X}}_l-\frac{1}{\varepsilon_l}(\mathbf{W}^{-2}\dot{\mathbf{A}}_l^H\dot{\mathbf{B}}_l-\mathbf{W}^{-2}\dot{\mathbf{C}}_l^H\dot{\mathbf{D}}_l),
\end{equation}
\begin{equation}
P_\Omega(\dot{\mathbf{X}}_*)=P_{\Omega^C}(\dot{\mathbf{X}}_*)+P_\Omega(\dot{\mathbf{M}}),
\end{equation}
where $lim_{k\rightarrow\infty}\dot{\mathbf{\mathbb{X}}}_k=\dot{\mathbf{X}}_*$, $\frac{1}{\varepsilon_k}$ is a positive step size.
\end{theorem}

Proof of Theorem \ref{theorem4} can be found in the Appendix \ref{A2} .

According to Theorem \ref{theorem4}, $\dot{\mathbf{X}}_{l+1}$ can be calculated directly by:
\begin{equation}
\dot{\mathbf{X}}_{l+1}=
\dot{\mathbf{X}}_l-\frac{1}{\varepsilon_l}(\mathbf{W}^{-2}\dot{\mathbf{A}}_l^H\dot{\mathbf{B}}_l-\mathbf{W}^{-2}\dot{\mathbf{C}}_l^H\dot{\mathbf{D}}_l).
\label{model33}
\end{equation}
\begin{equation}
P_\Omega(\dot{\mathbf{X}}_{l+1})=P_{\Omega^C}(\dot{\mathbf{X}}_{l+1})+P_\Omega(\dot{\mathbf{M}}),
\label{model34}
\end{equation}
Now, the weighted matrix is denoted as $\tilde{\mathbf{W}}=\mathbf{W}^{-2}=diag(\tilde{w}_i\cdots \tilde{w}_M)$, if $n_i\leq n_j$, the corresponding $\tilde{w}_i\geq \tilde{w}_j\geq1$, $n_i$ is the observed number of the i\emph{th} row. Besides, the update of $\dot{\mathbf{X}}_{l+1}$ is the solution of the following minimization problem which can be regard as a modification of (\ref{model21}) when searching the optimization by gradient descent method:

\begin{equation}\label{model35}
\begin{aligned}
&\min\limits_{\dot{\mathbf{X}}}
|tr(\dot{\mathbf{A}}\tilde{\mathbf{W}}\dot{\mathbf{X}}\dot{\mathbf{B}}^H)|-
|tr(\dot{\mathbf{C}}\tilde{\mathbf{W}}\dot{\mathbf{X}}\dot{\mathbf{D}}^H)|   \qquad   \\& \text{s.t.}  \quad P_\Omega(\dot{\mathbf{X}}-\dot{\mathbf{M}})=0,
\end{aligned}
\end{equation}
where $\dot{\mathbf{A}},\dot{\mathbf{B}},\dot{\mathbf{C}}$, and $\dot{\mathbf{D}}$ can be obtained analogously by (\ref{model17}), (\ref{model18}), (\ref{model19}), and (\ref{model20}), separately.
Let the initial $\dot{\mathbf{X}}_1=P_\Omega(\dot{\mathbf{M}})$, then we can solve it by one-step gradient descent strategy as follows:
\begin{equation}\label{model36}
\dot{\mathbf{X}}_{k+1}=
\dot{\mathbf{X}}_k-\frac{1}{\varepsilon_k}(\tilde{\mathbf{W}}\dot{\mathbf{A}}_k^H\dot{\mathbf{B}}_k
-\tilde{\mathbf{W}}\dot{\mathbf{C}}_k^H\dot{\mathbf{D}}_k).
\end{equation}
\begin{equation}\label{model37}
P_\Omega(\dot{\mathbf{X}}_{k+1})=P_{\Omega^C}(\dot{\mathbf{X}}_{k+1})+P_\Omega(\dot{\mathbf{M}}),
\end{equation}
where $\frac{1}{\varepsilon_k}$ is produced by:
\begin{equation}\label{model38}
\varepsilon_{k+1}=\rho\varepsilon_k,\qquad \varepsilon_1>0, \rho>1, k=1, 2, \cdots.
\end{equation}

The whole process named WQTNN and be summarized in Table \ref{table2}.

\begin{table}[htbp]
\caption{The one-step WQTNN algorithm}
\hrule
\label{table2}
\begin{algorithmic}[1]
\REQUIRE   the incomplete quaternion matrix data $\dot{\mathbf{M}}\in\mathbb{H}^{m\times n}$, the observed set $\Omega$, $\rho$, $\varepsilon_{max}$, weighted real matrix $\tilde{\mathbf{W}}$, and the tolerance $\epsilon_0$.
\STATE \textbf{Initial} $\dot{\mathbf{X}}_1=P_\Omega(\dot{\mathbf{M}})$, $\varepsilon_0$.
\STATE \textbf{Repeat}
\STATE \quad  Given $\dot{\mathbf{X}}_{k}$\\
 \qquad  \qquad \qquad  $[\dot{\mathbf{U}}_{k},\mathbf{\Sigma}_{k},\dot{\mathbf{V}}_{k}]=QSVD(\dot{\mathbf{X}}_{k})$
\STATE  \quad where $\dot{\mathbf{U}}_{k}=(\mathbf{u}_{1},\cdots\mathbf{u}_{m})\in\mathbb{H}^{m\times m}$,\\
       \qquad \quad\quad $\dot{\mathbf{V}}_{k}=(\mathbf{v}_{1},\cdots\mathbf{v}_{n})\in\mathbb{H}^{n\times n}$.
\STATE   \quad Computing $\dot{\mathbf{A}}_k=\mathbf{U}_{k}^H$ and $\dot{\mathbf{B}}_k=\mathbf{V}_{k}^H$ ,
\STATE   \quad Computing $\dot{\mathbf{C}}_k=(\mathbf{u}_{1},\cdots\mathbf{u}_{r})^H\in\mathbb{H}^{r\times m}$ and \\ \qquad\qquad\quad\quad$\dot{\mathbf{D}}_k=(\mathbf{v}_{1},\cdots\mathbf{v}_{r})^H\in\mathbb{H}^{r\times n}$,
\STATE \quad Computing $\dot{\mathbf{X}}_{k+1}=
\dot{\mathbf{X}}_k-\frac{1}{\varepsilon_k}(\tilde{\mathbf{W}}\dot{\mathbf{A}}_k^H\dot{\mathbf{B}}_k
-\tilde{\mathbf{W}}\dot{\mathbf{C}}_k^H\dot{\mathbf{D}}_k)$,
\STATE \quad   $P_\Omega(\dot{\mathbf{X}}_{k+1})=P_{\Omega^C}(\dot{\mathbf{X}}_{k+1})+P_\Omega(\dot{\mathbf{M}})$,
\STATE \quad Computing  $  \varepsilon_{k+1}=\min(\rho\varepsilon_k,\varepsilon_{max})$.
\STATE $k\longleftarrow k+1$
\STATE \textbf{Until convergence} $\|\dot{\mathbf{X}}_{k+1}-\dot{\mathbf{X}}_{k}\|_F /\|\dot{\mathbf{M}}\|_F\leq \epsilon_0$
\ENSURE  the recovered quaternion matrix $\dot{\mathbf{X}}_{rec}$.
\end{algorithmic}
\hrule
\end{table}

The weighted matrix in WQTNN method is used to control the rows' recovered priority and accuracy of the target missing quaternion matrix. Besides, the weighted matrix control two quaternion matrices $\dot{\mathbf{A}}_k^H\dot{\mathbf{B}}_k$ and $\dot{\mathbf{C}}_k^H\dot{\mathbf{D}}_k$ equally at the process of optimizing $\dot{\mathbf{X}}_{k+1}$. Further, to improve the accuracy of the recovery, we change weighted matrix in model (\ref{model35}) to control different weights for $\dot{\mathbf{A}}_k^H\dot{\mathbf{B}}_k$ and $\dot{\mathbf{C}}_k^H\dot{\mathbf{D}}_k$ respectively, named DWQTNN.

Being similar to the derivative process of WQTNN method, (\ref{model35}) can be modified for DWQTNN as follows:
\begin{equation}\label{model40}
\begin{aligned}
&\min\limits_{\dot{\mathbf{X}}}
|tr(\dot{\mathbf{A}}\mathbf{W}_1\dot{\mathbf{X}}\dot{\mathbf{B}}^H)|-
|tr(\dot{\mathbf{C}}\mathbf{W}_2\dot{\mathbf{X}}\dot{\mathbf{D}}^H)|   \qquad   \\& \text{s.t.}  \quad P_\Omega(\dot{\mathbf{X}}-\dot{\mathbf{M}})=0,
\end{aligned}
\end{equation}
where $\dot{\mathbf{A}},\dot{\mathbf{B}},\dot{\mathbf{C}}$, and $\dot{\mathbf{D}}$ can be obtained analogously by (\ref{model17}), (\ref{model18}), (\ref{model19}), and (\ref{model20}), separately. The weighted matrices $\mathbf{W}_1=diag(\tilde{w}_1\cdots \tilde{w}_M)$ and $\mathbf{W}_2=diag(\hat{w}_1\cdots \hat{w}_M)$ have the same function form like $\tilde{\mathbf{W}}$ in (\ref{model35}). Utilizing the number of observed entries ($m_i, i=1\cdots M$) in different rows, e.g.,if $m_i\leq m_j$, we set the corresponding $\hat{w}_i\geq \hat{w}_j\geq1$ for $\mathbf{W}_2$.

Given the initial $\dot{\mathbf{X}}_1=P_\Omega(\dot{\mathbf{M}})$, then we can also optimize (\ref{model40}) by one-step gradient descent strategy as follows:
\begin{equation}\label{model41}
\dot{\mathbf{X}}_{k+1}=
\dot{\mathbf{X}}_k-\frac{1}{\varepsilon_k}(\mathbf{W}_1\dot{\mathbf{A}}_k^H\dot{\mathbf{B}}_k
-\\ \mathbf{W}_2\dot{\mathbf{C}}_k^H\dot{\mathbf{D}}_k),
\end{equation}
\begin{equation}\label{model42}
P_\Omega(\dot{\mathbf{X}}_{k+1})=P_{\Omega^C}(\dot{\mathbf{X}}_{k+1})+P_\Omega(\dot{\mathbf{M}}).
\end{equation}

The requirements of convergence and the restriction of the step size $\frac{1}{\varepsilon_k}$ are given in the following theorem:

\begin{theorem}
\label{theorem5}
If $0 < \varepsilon_k < \varepsilon_{k+1}$, for $k=1, 2,\cdots $, $\lim_{k\rightarrow\infty}\frac{1}{\varepsilon_k}=0$.  the consequence $\dot{\mathbf{X}}_{k}$ produced by DWQTNN method will converge
with: $\|\dot{\mathbf{X}}_{N+1}-\dot{\mathbf{X}}_N\|_F\leq\epsilon$ and $N\geq1-\frac{\ln(\varepsilon_1\epsilon)-\ln(c)}{\ln(\rho)}$. Where $c=\|\mathbf{W}_1\|_Fm^{\frac{1}{2}}+\|\mathbf{W}_2\|_Fr^{\frac{1}{2}}$, m is the number of row of $\dot{\mathbf{X}}$, and $r$ is the truncated number.
\end{theorem}

Proof can be found in appendix \ref{A3}.
The  process of DWQTNN is summarized in Table \ref{table3}.

When the weight matrix $\mathbf{W}_2$  is same as $\mathbf{W}_1$ in DWQTNN algorithm, the DWQTNN algorithm would degrade into the WQTNN algorithm. To demonstrate the effectiveness of our method specifically, we give out connection with exiting works in quaternion domain.

\subsection{Connection With Exiting Works }
In this section, we explain the difference and the connection between our method and existing state-of-the-are quaternion-based methods \cite{8844978,9204671}.
\begin{itemize}
\item In \cite{8844978}, QNN is used to depict the low-rank structure of the quaternion matrix. This method needs to compute large-scale QSVD at every iteration, thus the computation cost is high. To avoid computing the QSVD of a large quaternion matrix in each step, the work in \cite{9204671} modelled the low-rank property by using low-rank quaternion matrix factorization. The calculation of two smaller size quaternion matrices' QSVD is also necessary for \cite{9204671}. Our method utilizes QTNN, in this way, we only need to compute QSVD a few times rather than at every iteration, and the main target of QTNN is to solve a subproblem (\ref{model6}).
\item Further, we improve the QTNNR by adding weighted matrices. The subsequently proposed DWQTNNR algorithm only need one-step to implement recover which replaces the QSVT that is another difference with the exiting quaternion-based method like \cite{8844978}. Comparing our method with other quaternion-based methods, we consider more information about the missing situations of the incomplete data, and the one-step DWQTNNR could be faster to get the inpainting results.
\end{itemize}
\begin{table}[htbp]
\caption{The DWQTNN algorithm}
\hrule
\label{table3}
\begin{algorithmic}[1]
\REQUIRE   the incomplete quaternion matrix data $\dot{\mathbf{M}}\in\mathbb{H}^{m\times n}$, the observed set $\Omega$, $\rho$, $\varepsilon_{max}$ , weighted real matrices $\mathbf{W}_1$ and $\mathbf{W}_2$, and the tolerance $\epsilon_0$.
\STATE \textbf{Initial} $\dot{\mathbf{X}}_1=P_\Omega(\dot{\mathbf{M}})$, $\varepsilon_0$.
\STATE \textbf{Repeat}
\STATE \quad \textbf{Step 1.} Given $\dot{\mathbf{X}}_{k}$\\
       \qquad  \qquad \qquad  $[\dot{\mathbf{U}}_{k},\mathbf{\Sigma}_{k},\dot{\mathbf{V}}_{k}]=QSVD(\dot{\mathbf{X}}_{k})$
\STATE \quad where $\dot{\mathbf{U}}_{k}=(\mathbf{u}_{1},\cdots\mathbf{u}_{m})\in\mathbb{H}^{m\times m}$,\\
       \qquad \quad\quad $\dot{\mathbf{V}}_{k}=(\mathbf{v}_{1},\cdots\mathbf{v}_{n})\in\mathbb{H}^{n\times n}$.
\STATE   \quad Computing $\dot{\mathbf{A}}_k=\mathbf{U}_{k}^H$ and $\dot{\mathbf{B}}_k=\mathbf{V}_{k}^H$ ,
\STATE   \quad Computing $\dot{\mathbf{C}}_k=(\mathbf{u}_{1},\cdots\mathbf{u}_{r})^H\in\mathbb{H}^{r\times m}$ and \\ \qquad\qquad\quad\quad$\dot{\mathbf{D}}_k=(\mathbf{v}_{1},\cdots\mathbf{v}_{r})^H\in\mathbb{H}^{r\times n}$,
\STATE \quad Computing $\dot{\mathbf{X}}_{k+1}=$\\
\qquad\qquad\quad\quad$\dot{\mathbf{X}}_k-\frac{1}{\varepsilon_k}(\mathbf{W}_1\dot{\mathbf{A}}_k^H\dot{\mathbf{B}}_k
                -\mathbf{W}_2\dot{\mathbf{C}}_k^H\dot{\mathbf{D}}_k)$,
\STATE \quad   $P_\Omega(\dot{\mathbf{X}}_{k+1})=P_{\Omega^C}(\dot{\mathbf{X}}_{k+1})+P_\Omega(\dot{\mathbf{M}})$,
\STATE \quad Computing $\varepsilon_{k+1}=\min(\rho\varepsilon_k,\varepsilon_{max})$.
\STATE $k\longleftarrow k+1$
\STATE \textbf{Until convergence} $\|\dot{\mathbf{X}}_{k+1}-\dot{\mathbf{X}}_{k}\|_F /\|\dot{\mathbf{M}}\|_F\leq \epsilon_0$
\ENSURE  the recovered quaternion matrix $\dot{\mathbf{X}}_{rec}$.
\end{algorithmic}
\hrule
\end{table}
\section{Experimental results}\label{Experimental results}
In this section, we perform some experiments to evaluate the performance of our method (QTNNR and DWQTNN). Comparative experiments are conducted by the following low-rank completion methods:
\begin{enumerate}
  \item \textbf{WNNM} \cite{gu2017weighted}: this method based on the weighted nuclear norm of the real matrix to implement the LRMC.
  \item \textbf{TNNR} \cite{6389682}: this method based on the truncated nuclear norm of the real matrix to implement the LRMC.
  \item \textbf{TNNR-WRE} \cite{liu2015truncated}: this method based on the truncated nuclear norm of the real matrix with weight residual error to implement the LRMC.
  \item \textbf{ETNNR-WRE} \cite{liu2015truncated}: this method is an extension model of TNNR-WRE method.
  \item \textbf{DWTNNR} \cite{DBLP:journals/corr/abs-1901-01711}: this method is an extension model of TNNR method by adding two weighted matrices in the optimization.
  \item \textbf{LRQA-2} \cite{8844978}: this method based on QSVT, replace the QNN by utilizing Laplace function to implement the LRQMC.
  \item \textbf{Q-DNN} \cite{9204671}: this method based on low-rank quaternion matrix factorization to implement the LRQMC.
\end{enumerate}

All the experiments are executed in Matlab R2018b, on a computer with a 1.60GHz CPU and 8GB memory based on Windows 10.

\textbf{Parameters setting:} For QTNNR, we let $\rho=1.25$, $\beta_{max}=10^7$ and $\beta_0=0.005$. The stopping criterion is $\|\dot{\mathbf{X}}_{k+1}-\dot{\mathbf{X}}_{k}\|_F /\|\dot{\mathbf{M}}\|_F\leq \epsilon_0$, where $\epsilon_0=0.001$. By reasons of the absence of prior knowledge to the number of truncated singular values, $r$ is tested from [1, 10] to choose an optimal value for each case manually. 

For DWQTNN, we let $\rho=1.2$, $\beta_{max}=10^7 $ and $\beta_0=0.0015$. The stopping criterion is $\|\dot{\mathbf{X}}_{k+1}-\dot{\mathbf{X}}_{k}\|_F /\|\dot{\mathbf{M}}\|_F\leq \epsilon_0$, where $\epsilon_0=0.0001$, $r$ is tested from [1, 20] to choose an optimal value for each case manually. Another committed operation in DWQTNN is to determine the weighted matrices $\mathbf{W}_1$ and $\mathbf{W}_2$, and these weighted matrices are determined by the number of observed entries in each row. Basing on the property of weighted matrix derived from (\ref{model33}), the weighted numerical values at the main diagonal of the weighted matrix $\mathbf{W}_1$ and $\mathbf{W}_2$  are given as follows separately:
\begin{equation}\label{model43}
   \tilde{w}_i=\theta_1(2-\frac{m_i^R}{m}),\quad i=1,2,\cdots M,
  \end{equation}
where $m_i^R$ is the number of the number of observed entries in the $i$\emph{th} row, and $\theta_1$ is used to control the weights.
\begin{equation}\label{model44}
   \hat{w}_i=\theta_2(2-\frac{m_i^R}{n}),\quad i=1,2,\cdots N,
  \end{equation}
where $m_i^R$ is the number of the number of observed entries in the $i$\emph{th} row, and $\theta_2$ is used to control the weights. When $\theta_1$ and $\theta_2$ setting to zero, it means that the weighted matrix degrade into the identity matrix. When $\theta_1=\theta_2$, it means that the DWQTNN method degrade into the WQTNN. Besides, we also can make the weighed matrices be placed on the right side of the residual error and the weights can be decided by the missing number of each column.  In the following experiment, we set $\theta_1=2$ and $\theta_2=1.5$ for DWQTNN. At the first experiment, we test the effect of the added weighted matrices by recovering the block missing color images and give out the comparison with other weighted-LRMC methods \cite{liu2015truncated, DBLP:journals/corr/abs-1901-01711}.

\textbf{Performance index setting:} The peak signal to noise rate (PSNR) \cite{liu2015truncated} and the structural similarity index (SSIM) \cite{1284395} are two frequently indexes adopted for measuring the quality of the recovered results, and we compare the performance by measuring these two indexes. The \textbf{bold} fonts represent the best results, and the \underline{underline} ones denote the second-best performance at the following experiments.

\begin{figure}
 \centering
 \includegraphics[width=100mm]{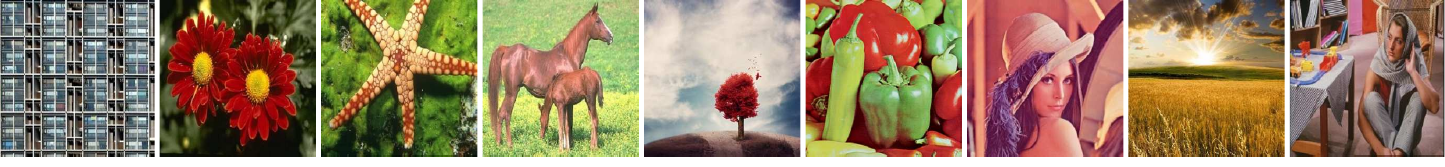}
 \caption{The $9$ color images (from left to right, Image(1) $300\times 300\times 3$,
  Image(2)  $300\times 300\times 3$, Image(3)  $300\times 300\times 3$, Image(4)  $300\times 300\times 3$, Image(5)  $300\times 300\times 3$, Image(6)  $300\times 300\times 3$, Image(7)  $300\times 300\times 3$,  Image(8)  $300\times 300\times 3$), and Image(9) $300\times 300\times 3$).}
 \label{Ytu}
\end{figure}

The compared methods' parameters are set as the experimental settings reported in their papers individually \cite{gu2017weighted, 6389682,liu2015truncated,DBLP:journals/corr/abs-1901-01711,8844978,9204671}. 9 frequently used color images are selected as the test samples that are shown in Fig. \ref{Ytu}. Because image (1) has many similar blocks, it is used to test the efficiency of weighted matrix-like in \cite{liu2015truncated} by comparing block missing completion.

\subsection{The effectiveness of the weighted matrices}
In this experiment, we mask image(1) and image(8) with missing blocks to illustrate the effectiveness of the added weighted matrices.

\begin{figure*}
\centering
\subfigure[Block missing]{
\includegraphics[width=3cm]{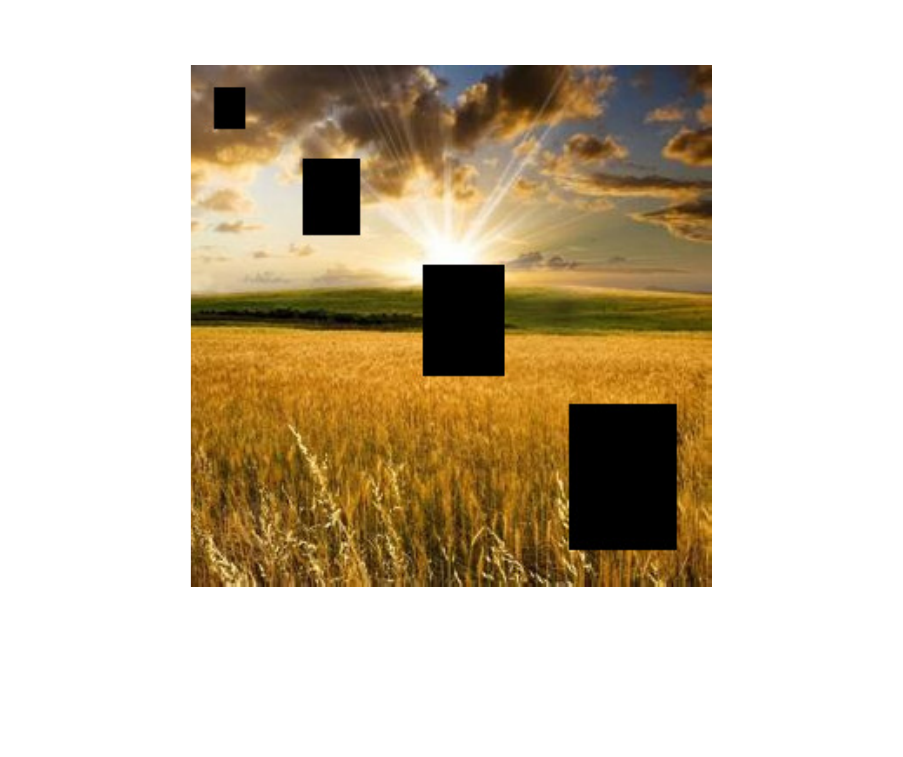}
}
\hspace{-1cm}
\subfigure[15\emph{th} step]{
\includegraphics[width=3cm]{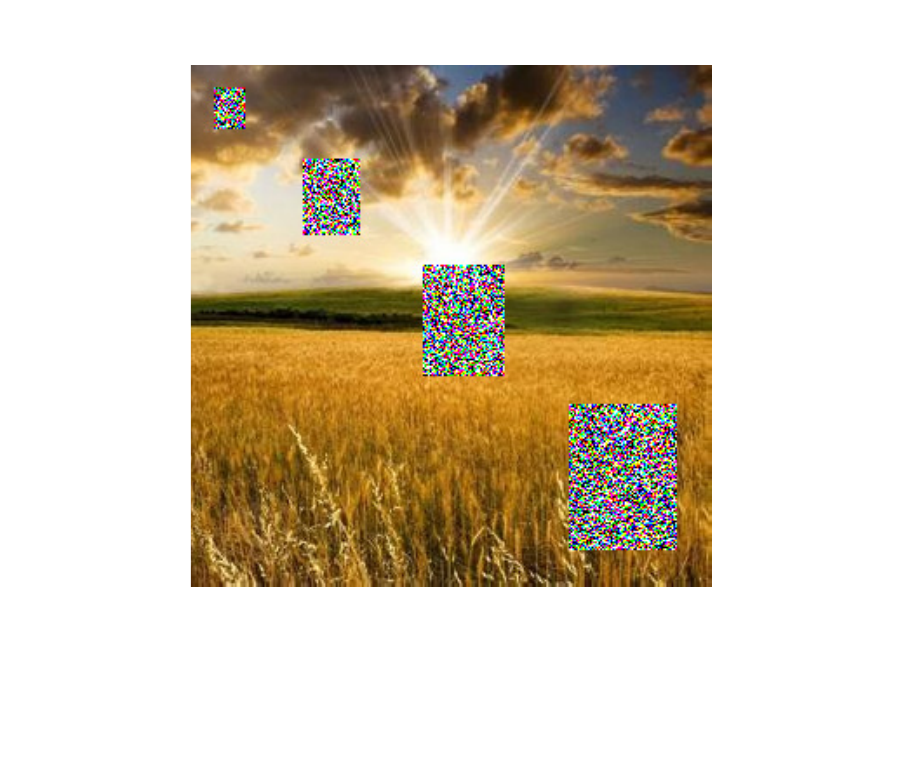}
}
\hspace{-1cm}
\subfigure[25\emph{th} step]{
\includegraphics[width=3cm]{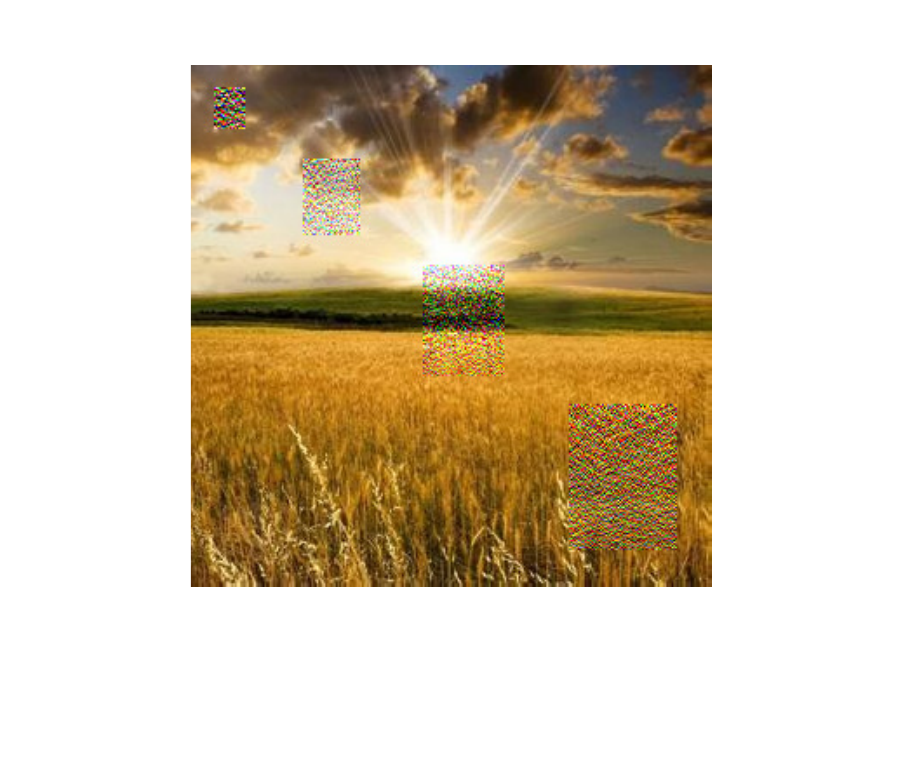}
}
\hspace{-1cm}
\subfigure[30\emph{th} step]{
\includegraphics[width=3cm]{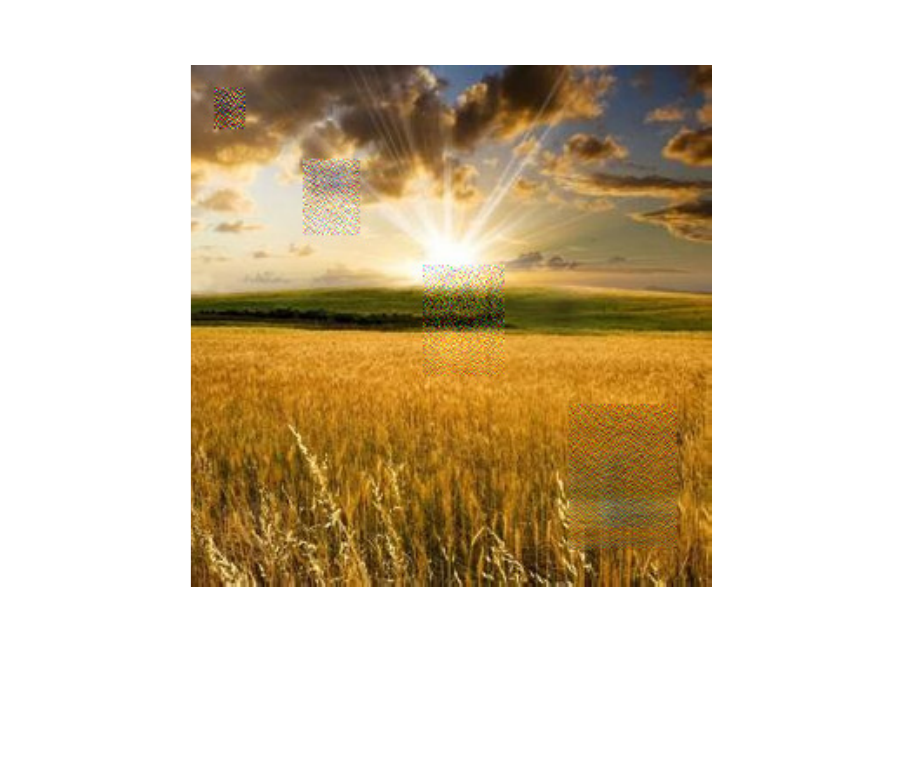}
}
\hspace{-1cm}
\subfigure[36\emph{th} step (convergence)]{
\includegraphics[width=3cm]{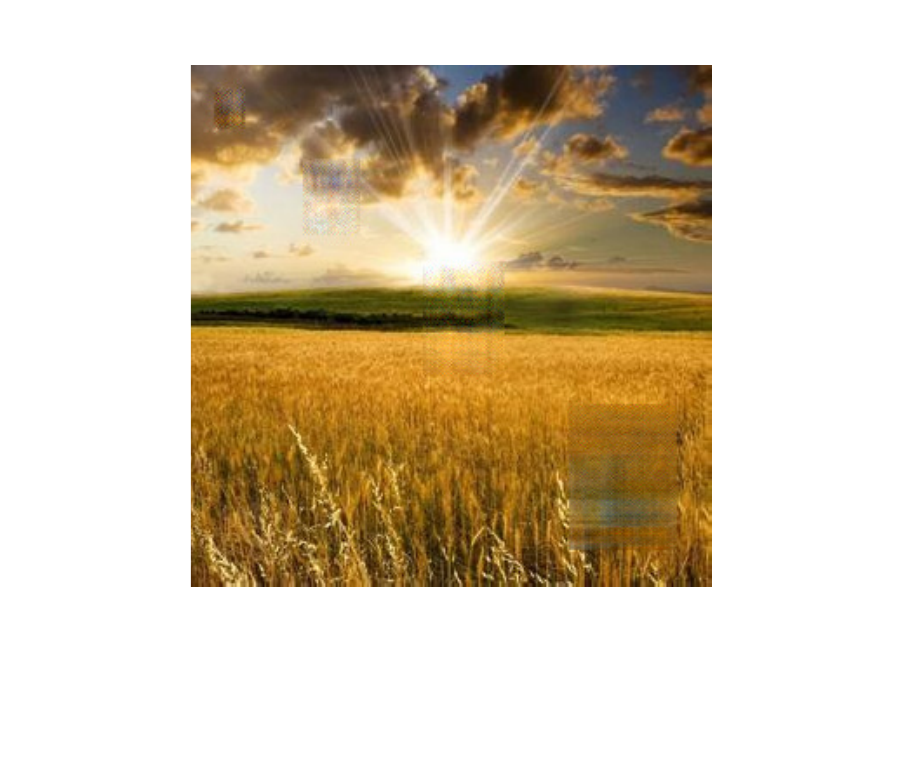}
}
\caption{ The process of recovering incomplete image (a) from  the 15\emph{th} step (b), the 25\emph{th} (c), the 30\emph{th} (d) to the 36\emph{th} step (e), and (e) is the recovered image. }
 \label{E1}
\end{figure*}

Fig. \ref{E1} illustrates the process of recovering image(8) at the 0\emph{th}, 15\emph{th}, 25\emph{th}, 39\emph{th} iterations,  respectively, and the final result (at the 36\emph{th} iteration). The smaller missing blocks correspond to smaller weights in the main diagonal of the weighted matrix, so that the upper part of the missing image can be recovered more prior and accurate.

\begin{figure*}
\centering
\subfigure[Original]{
\includegraphics[width=3cm]{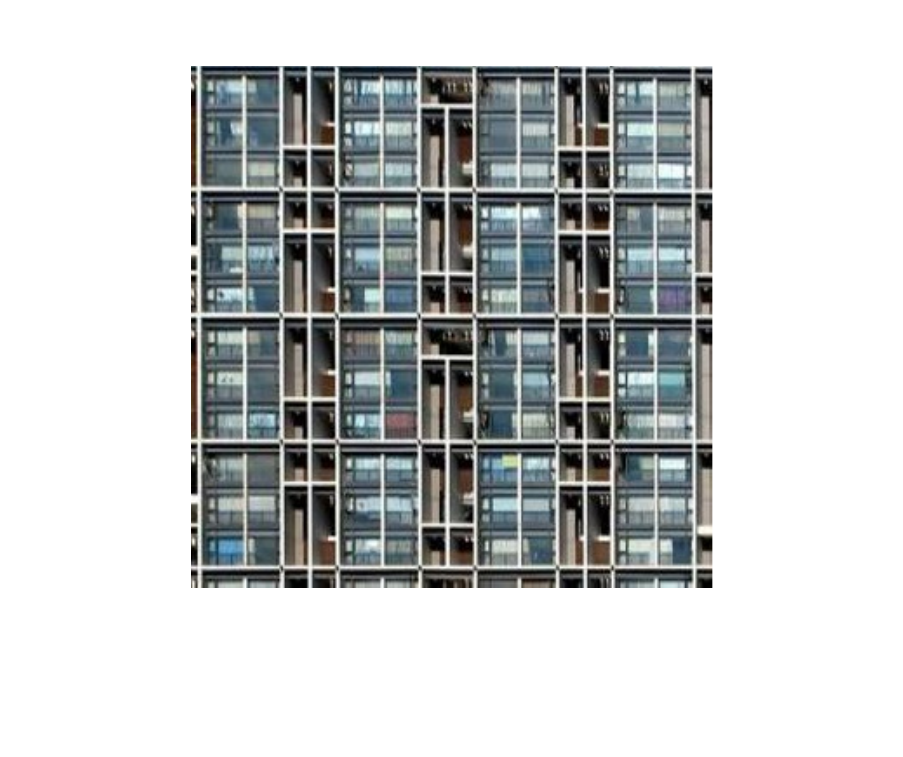}
}
\hspace{-1cm}
\subfigure[Square Missing]{
\includegraphics[width=3cm]{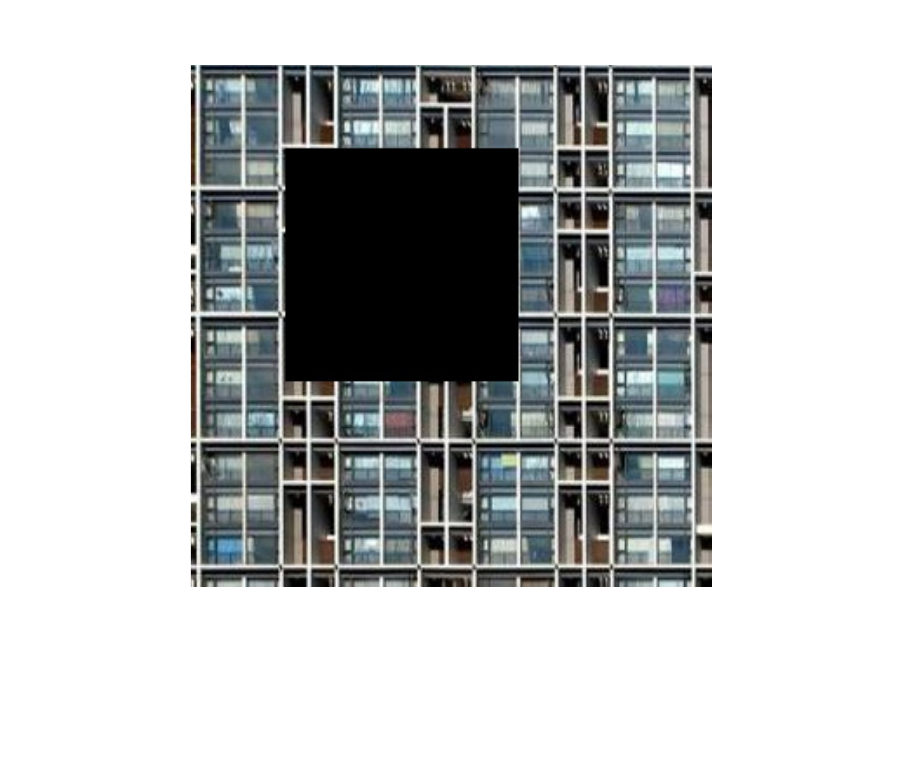}
}
\hspace{-1cm}
\subfigure[ETNNR-WRE]{
\includegraphics[width=3cm]{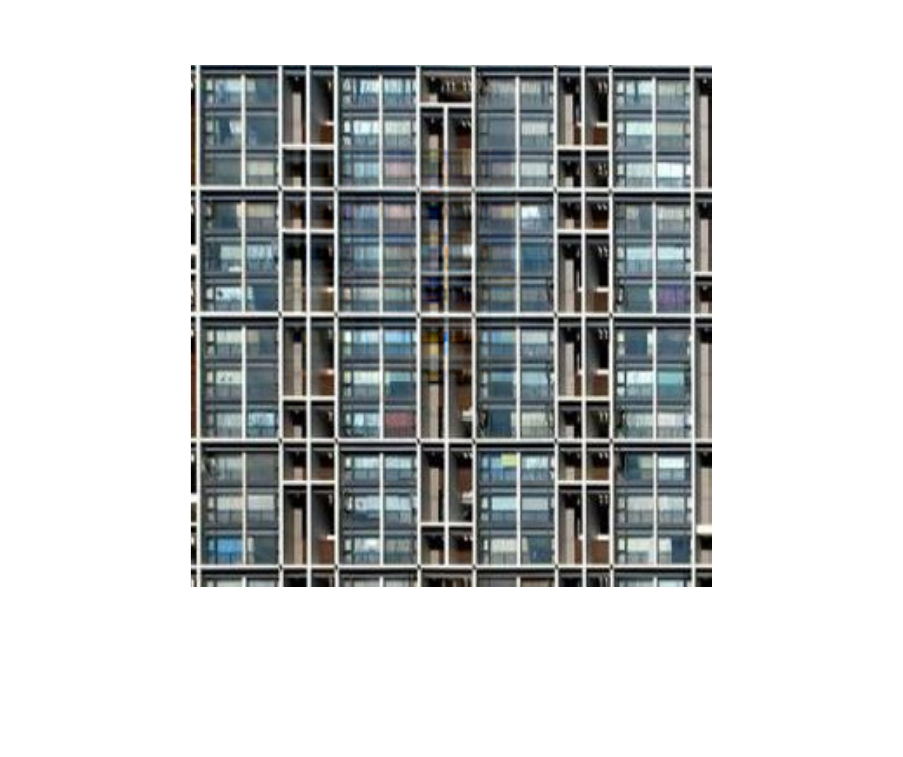}
}
\hspace{-1cm}
\subfigure[DWTNNR]{
\includegraphics[width=3cm]{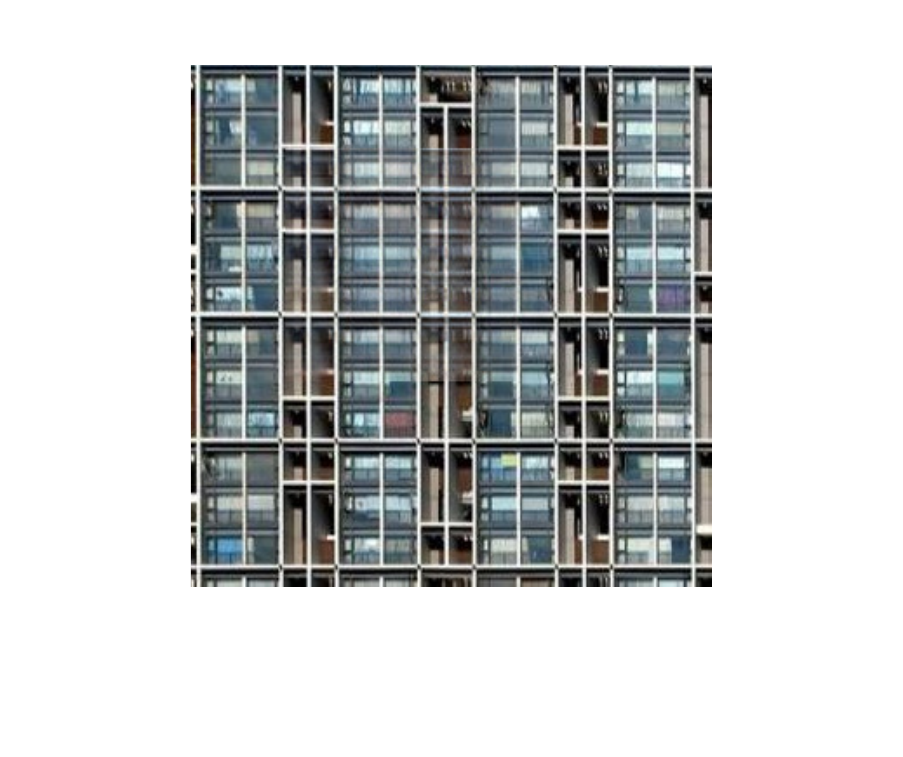}
}
\hspace{-1cm}
\subfigure[DWQTNN]{
\includegraphics[width=3cm]{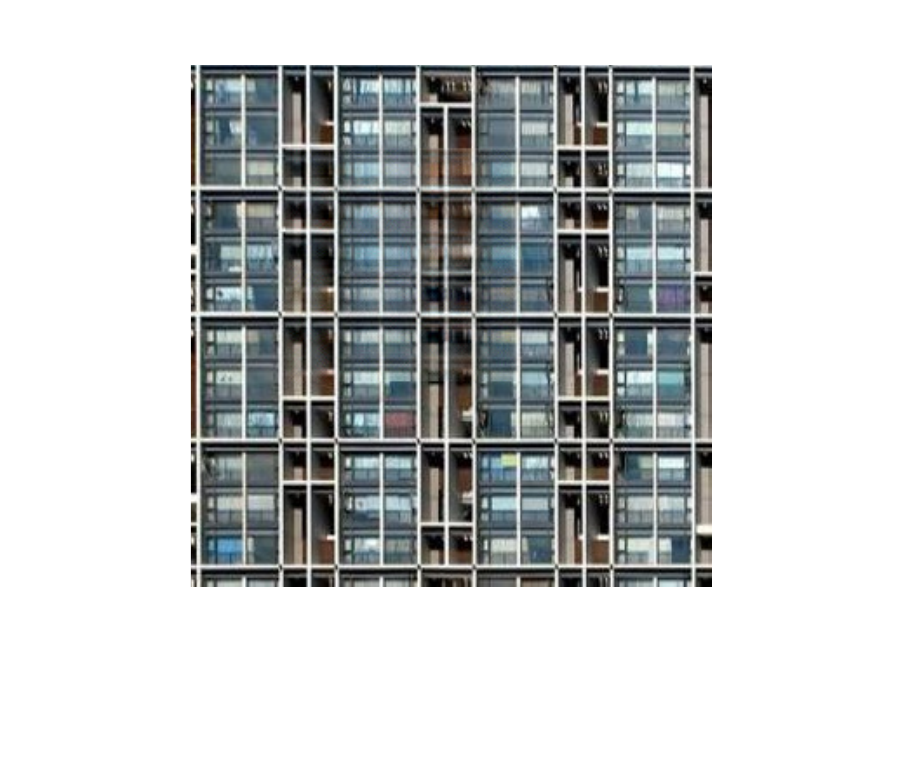}
}
\caption{The comparison of recovering incomplete image (b) by ETNNR-WRE  (c), PSNR = 24.012 (Time = 6.951s); DWTNNR (d), PSNR = 23.976 (Time = 10.365s); and DWQTNN (e), PSNR=\textbf{24.501} (Time = 10.850s).}
\label{E2}
\end{figure*}

Fig. \ref{E2} compares our method with other two weighted algorithms in LRTC (ETNNR-WRE and DWTNNR) by recovering image(1) which has many similar structures that is corroborative evidence for the effectiveness of weighted matrices.

\begin{figure}[htbp]
\centering
\subfigure[]{
\includegraphics[width=3cm]{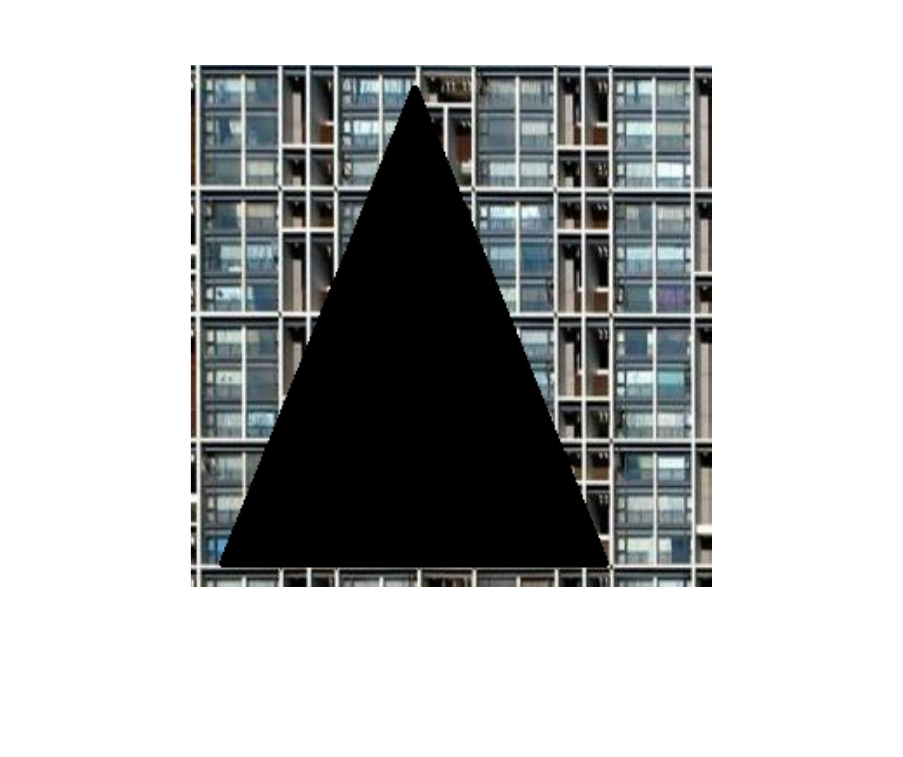}
}
\vspace{3pt}\hspace{-1cm}
\subfigure[ETNNR-WRE]{
\includegraphics[width=3cm]{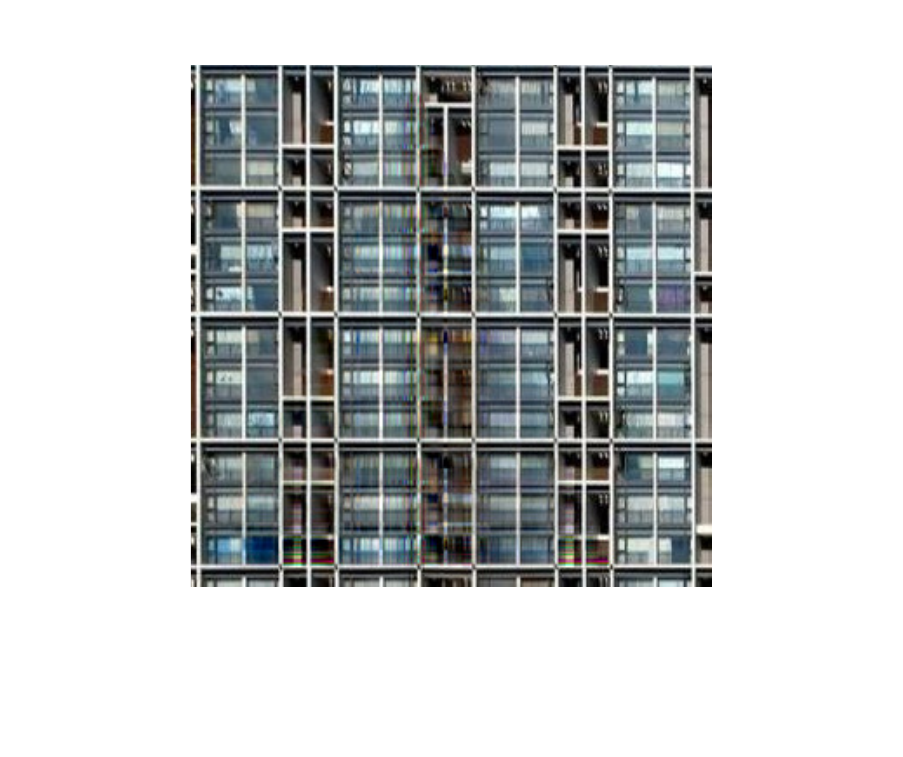}
}
\vspace{3pt}\hspace{-1cm}
\subfigure[DWTNNR]{
\includegraphics[width=3cm]{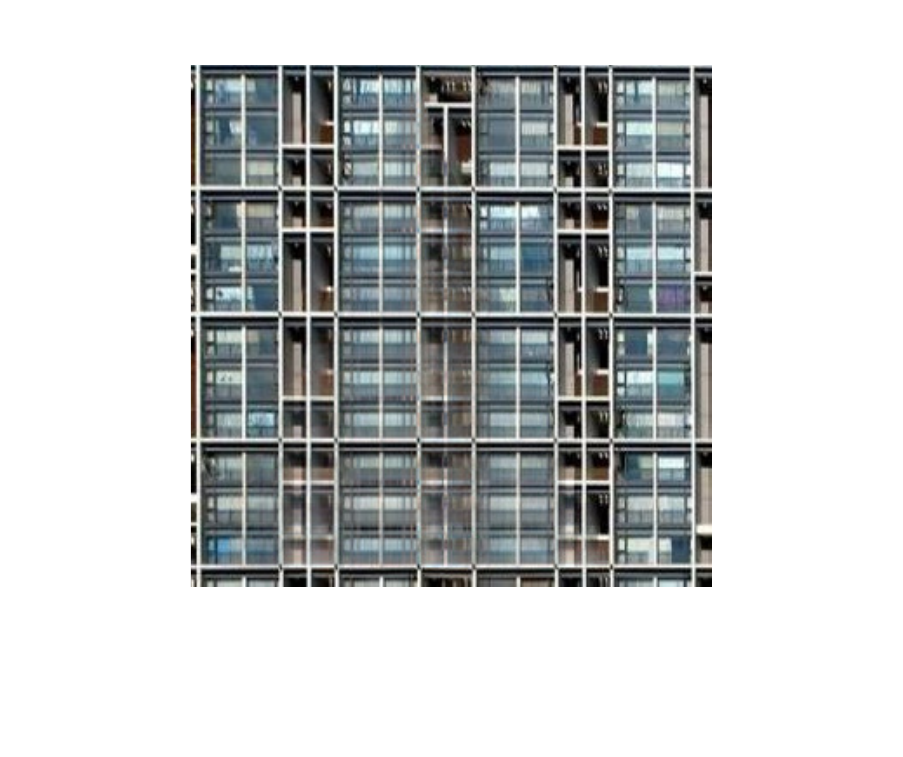}
}
\vspace{3pt}\hspace{-1cm}
\subfigure[DWQTNN]{
\includegraphics[width=3cm]{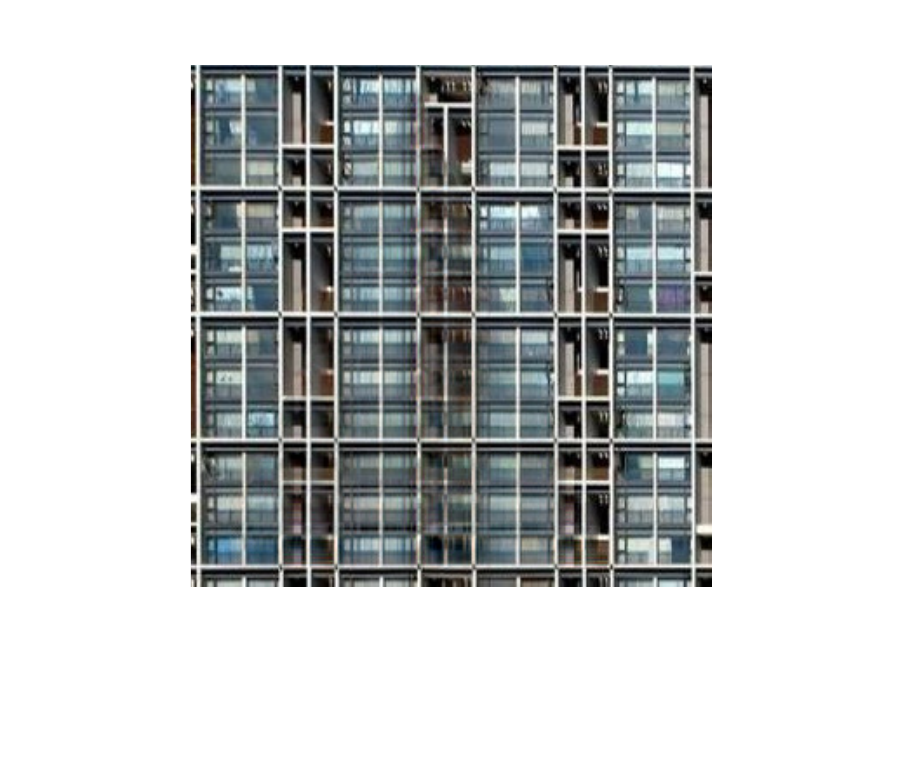}
}
\
\subfigure[]{
\includegraphics[width=3cm]{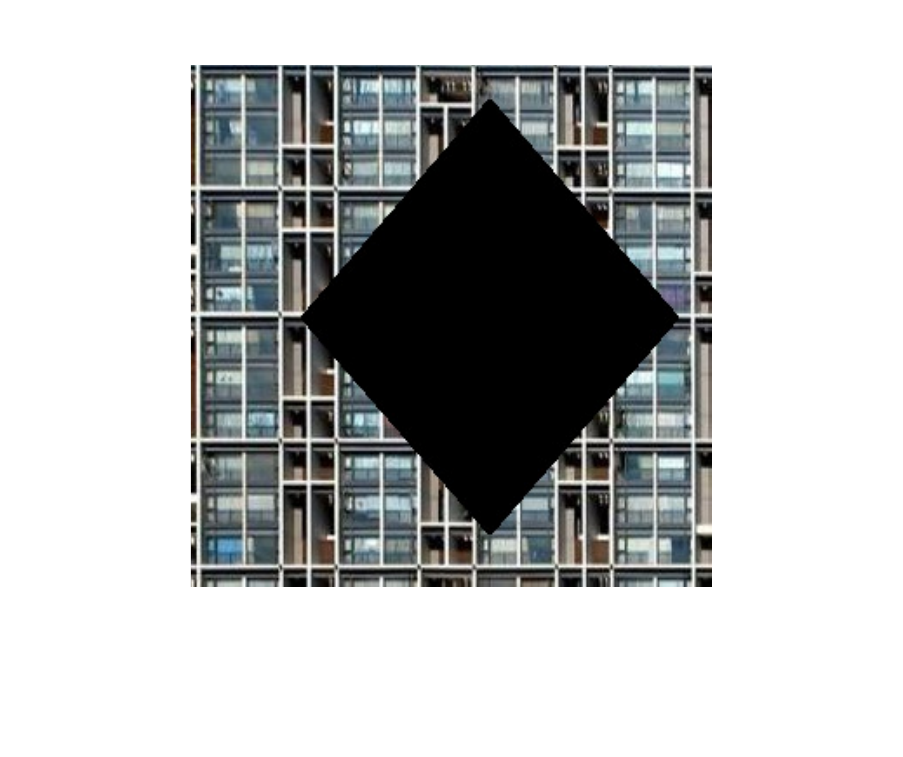}
}
\vspace{3pt}\hspace{-1cm}
\subfigure[ETNNR-WRE]{
\includegraphics[width=3cm]{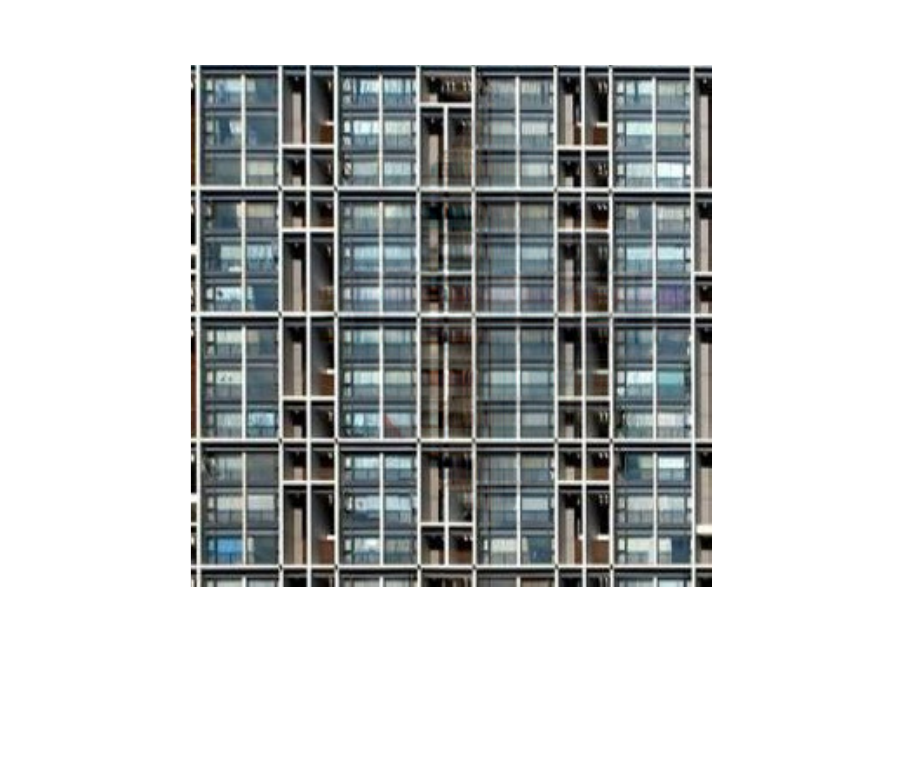}
}
\vspace{3pt}\hspace{-1cm}
\subfigure[DWTNNR]{
\includegraphics[width=3cm]{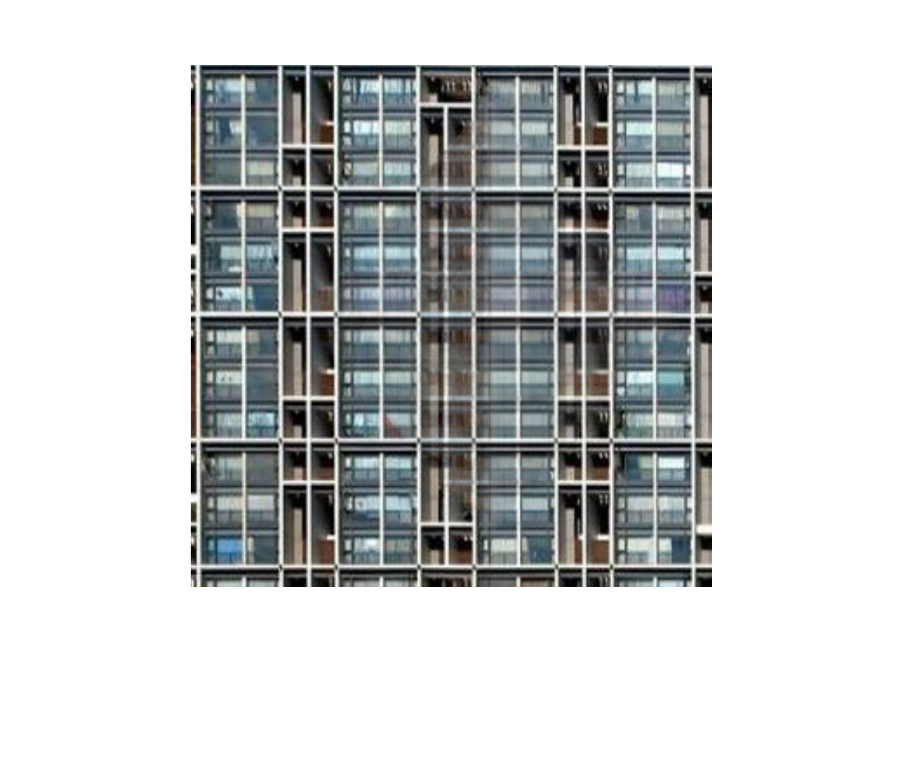}
}
\vspace{3pt}\hspace{-1cm}
\subfigure[DWQTNN]{
\includegraphics[width=3cm]{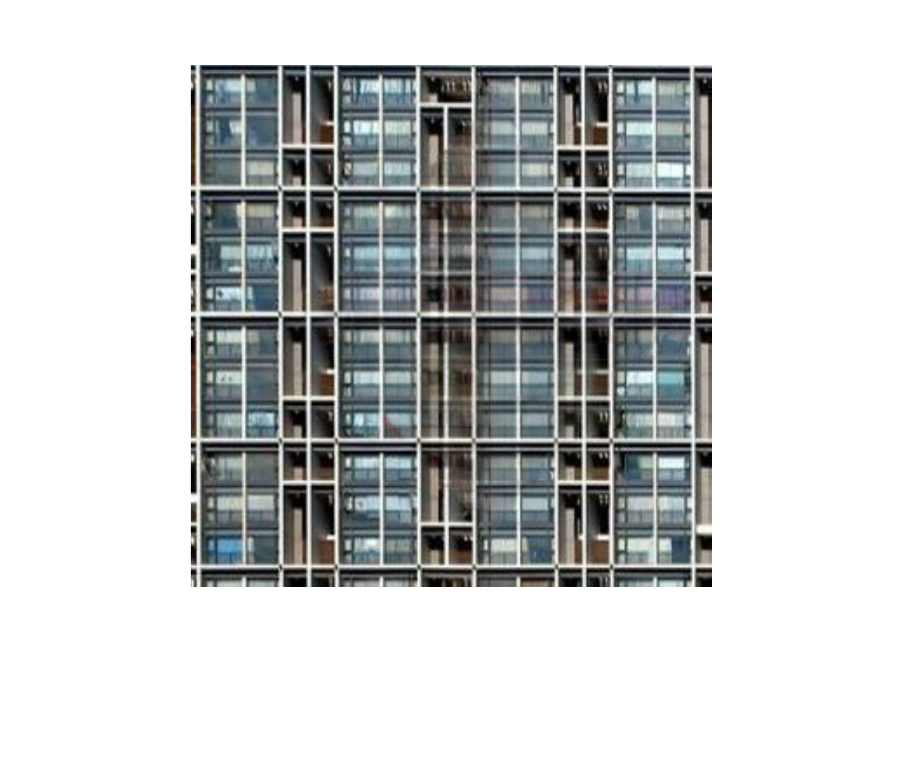}
}
\caption{The comparison of recovering incomplete image (a)  and (e) by  three weighted methods (ETNNR-WRE, DWTNNR, DWQTNN)}.
\label{E3}
\end{figure}

\begin{figure}[htbp]
\centering
\subfigure[]{
\includegraphics[width=3.5cm]{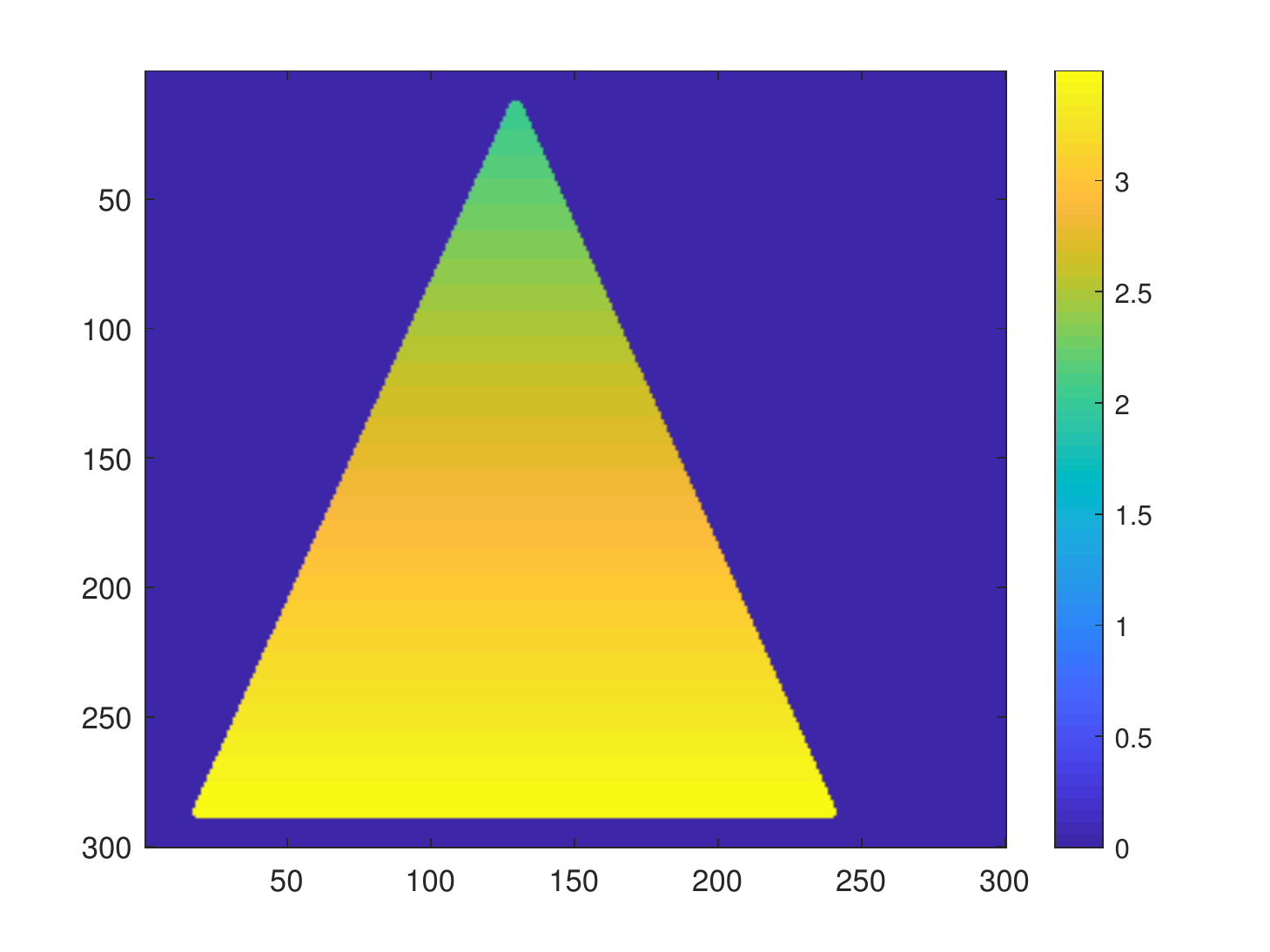}
}
\subfigure[]{
\includegraphics[width=3.5cm]{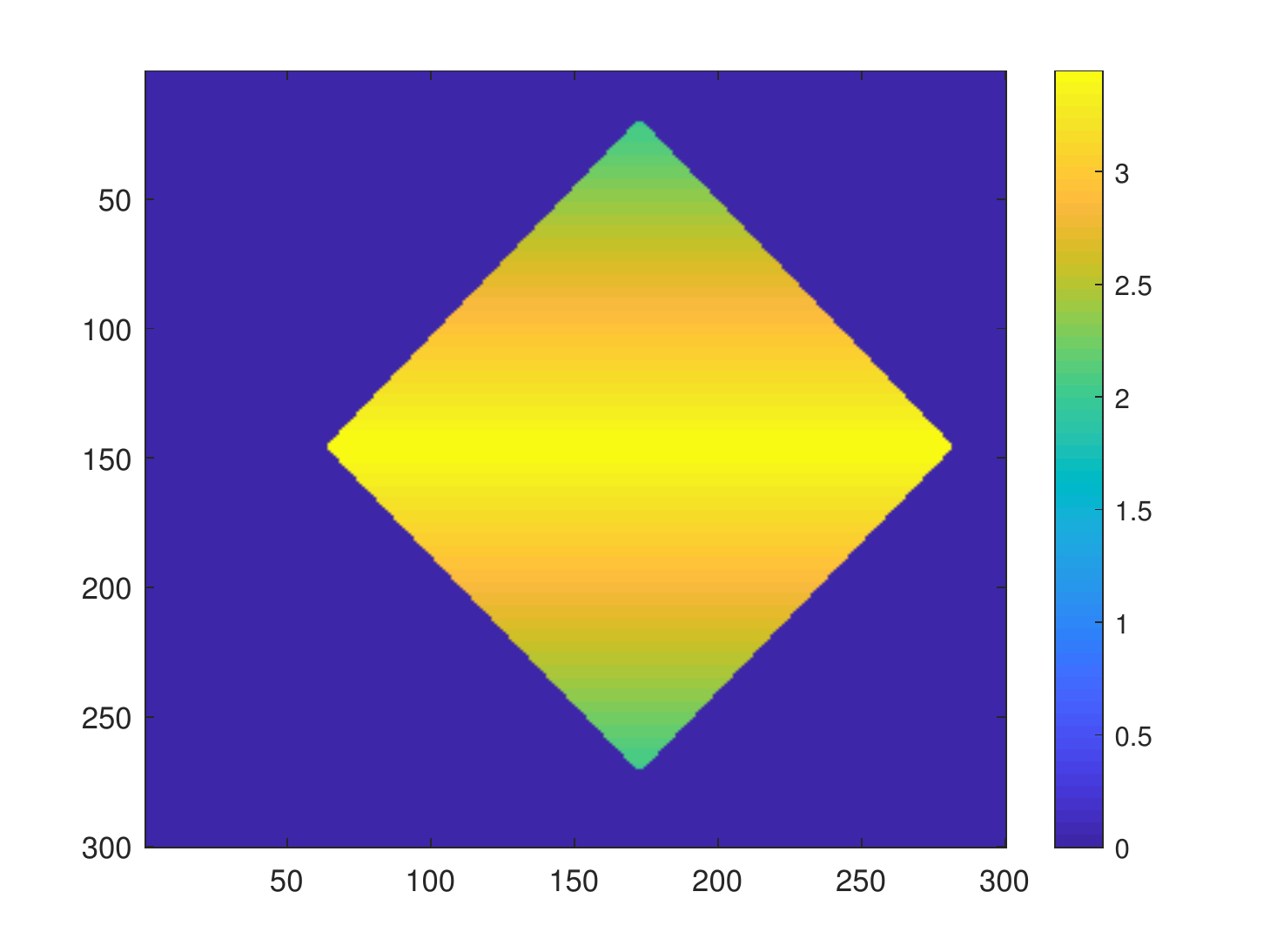}
}
\caption{The visualized weights in DWQTNN ($\mathbf{W}_1$) for Fig. \ref{E3} (a) and Fig. \ref{E3} (e).}
\label{E3.1}
\end{figure}

Fig. \ref{E3} displays the results of recovering another two different block missing forms that are masked in the image(1) to demonstrate the effectiveness of weighted matrices. The missing block in Fig. \ref{E3} (a) is an equilateral triangle block, it means that the number of observed entries in the rows of image(1) is restricted to decrease from the top to the bottom progressively, accordingly, the weights in the main diagonal of weighed matrices will increase progressively. The weighed matrix for the missing diamond block in Fig. \ref{E3} (e) is in analogy with the triangle block. The visualized weights are shown in Fig. \ref{E3.1}, and the corresponding index performance is shown in Table. \ref{E3i}.

From the above experiments about the weighted matrix, we can conclude that \emph{a.} The operation of adding the weights by formulation (\ref{model43}) and (\ref{model44}) decided by the observed entries in each row. If more observed entries in this row, it would be recovered with higher accuracy. \emph{b.} Our weighted method is controlled by the number of observed entries directly, and the ETNNR-WRE is to segment the weights by utilizing the missing structure. Besides, DWQTNN is based on the quaternion matrix such that RGB channels can be handled together, and we can utilize more information about the internal structure of the missing image.

\begin{table}
\caption{Corresponding PSNR and time of recovery Fig. \ref{E3} (a) and (e) by three weighted methods.}
\scalebox{0.75}{\begin{tabular}{cccc}
\hline
Method & ETNNR-WRE (b) & DWTNNR (c) & DWQTNN (d)\\
\hline
PSNR(Time(s)) & 18.665 (7.853s) & 18.970 (10.905s) & \textbf{19.267} (11.993s)\\
\hline
\hline
Method & ETNNR-WRE (f) & DWTNNR (g) & DWQTNN (h)\\
\hline
PSNR(Time(s)) & 21.067 (7.701s) & 20.938 (11.971s) & \textbf{21.412} (11.708s)\\
\hline
\end{tabular}\label{E3i}}
\end{table}

\subsection{Real Visual Data Completion}
In the image process, recovering missing data is a necessary target. In this section, we give out the results of completing images in Fig. \ref{Ytu} from image(2) to image(9). For tested images, at first, we load color images as a whole. Then, we randomly and consistently choose elements that are smaller than \emph{p}. Lastly, we let these positions where these be chosen elements in construct the set $\Omega$. The \emph{p} is named missing rate, and $0<p<1$ . The value of \emph{p} is larger, the missing rate is higher, and vice versa. We employ quantitative quality indexes, including  PSNR and SSIM. Besides, we compared the time consumption by these methods to recover eight images.

In Table \ref{E4}, we display the random missing image recover results with $\emph{p}=0.5$ by comparing the PSNR, SSIM values and the time consuming of our QTNN and DWQTNN  methods (setting $\theta_1=2$ and $\theta_2=1.5$) and other compared methods.

In Table \ref{E5} and Table \ref{E6}, we display the random missing image recover results with $\emph{p}=0.65$ and $\emph{p}=0.75$, respectively by comparing the PSNR, SSIM values and the time consuming of our QTNN and DWQTNN  methods with other two quaternion-based methods. The matrix-based methods would not get better performance than quaternion-based methods when the missing rate is high \cite{8844978}, \cite{9204671}, so we only concentrate on the comparison of quaternion-methods. The corresponding visualized comparison are in Fig. \ref{E0.65} and Fig. \ref{E0.75}.

\begin{figure}[htbp]
 \centering
 \includegraphics[width=80mm]{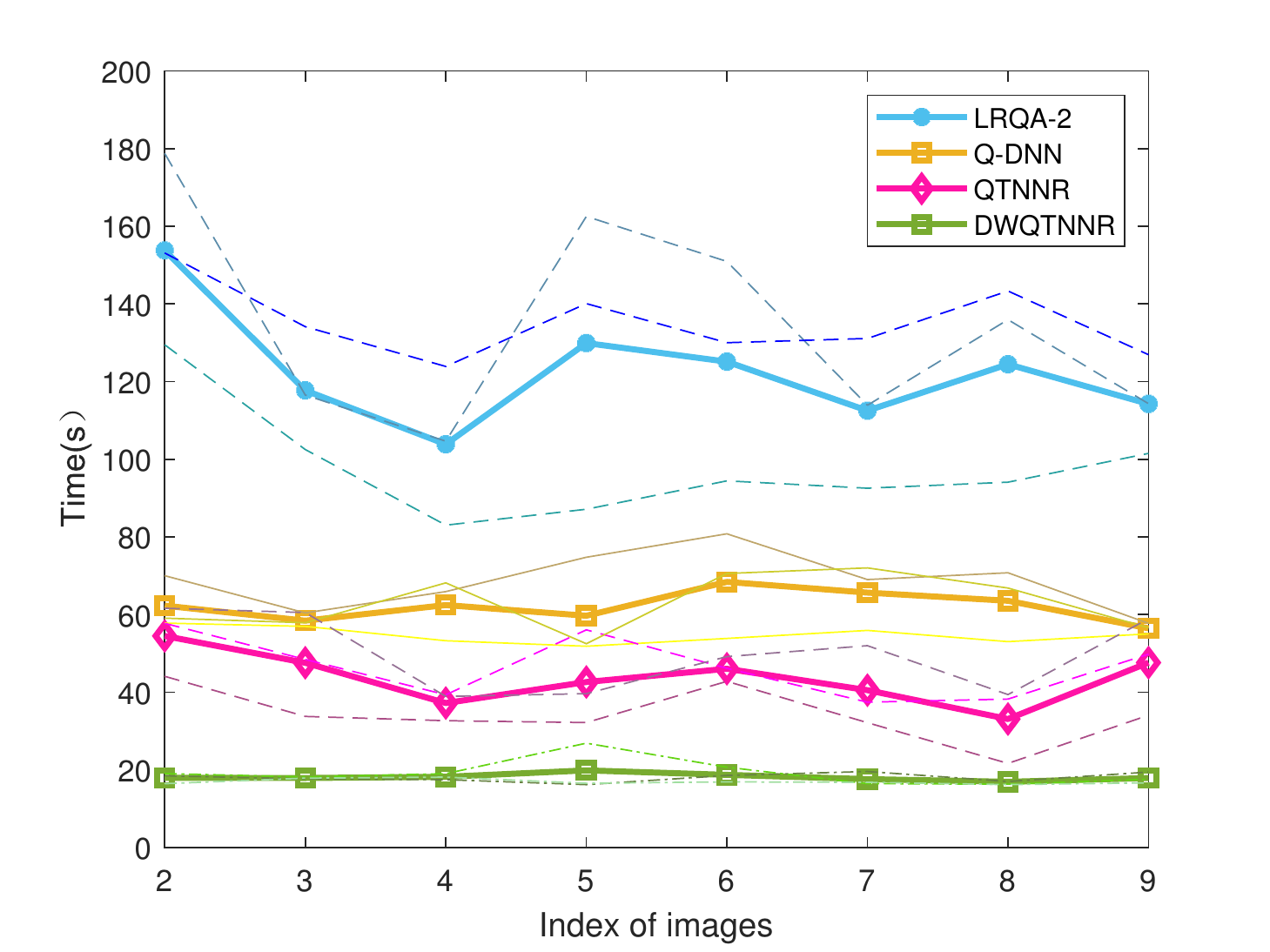}
 \caption{The time consumption by four methods on eight images with different missing values.}
\label{E7}
\end{figure}

Fig. \ref{E7} demonstrates the time consumption of all quaternion-based methods for recovering the testing color images with different missing rates that are we text in the experiments above. The fine lines are the time consumption of recovering missing images with \emph{p}=0.5, \emph{p}=0.65 and \emph{p}=0.75, separately. The bond lines are the average elapsed time for getting recover results by operating each method.

\renewcommand{\arraystretch}{1.5}
\begin{table*}
  \centering
  \fontsize{6.5}{8}\selectfont
  \begin{threeparttable}
  \caption{QUANTITATIVE EVALUATION (PSNR/SSIM and TIME Consumption(s)) OF DIFFERENT COLOR IMAGE COMPLETION ALGORITHMS}
  \scalebox{0.7}{
    \begin{tabular}{cccccccccc}
    \toprule
    \multirow{2}{*}{Method}&
    \multicolumn{5}{c}{Matrix-based}&\multicolumn{4}{c}{Quaternion-based}\cr
    \cmidrule(lr){2-6} \cmidrule(lr){7-10}
    &WNNM \cite{gu2017weighted}&TNNR \cite{6389682}&TNNR-WRE \cite{liu2015truncated}&ETNNR-WRE \cite{liu2015truncated}&DWTNNR  \cite{DBLP:journals/corr/abs-1901-01711}&LRQA-2 \cite{8844978}&Q-DNN \cite{9204671}&QTNN [ours]&DWQTNN [ours]\cr
    \midrule
    \multirow{2}*{Image(2)}&25.638/0.940&27.364/0.941&27.060/0.965&\underline{27.454/0.966}&26.670/0.961&27.430/0.965&26.791/0.955&\textbf{27.744/0.967}&27.166/\underline{0.966}\\
    \cline{2-10}
   & 83.441 & 41.232 & 13.334 &8.430 & 10.222 & 129.487 & 57.803 &44.100&18.483\\
   \cline{1-10}
    \multirow{2}*{Image(3)}&21.529/0.886&23.901/0.929&23.794/0.939&23.755/\underline{0.941}&23.153/0.936&23.881/0.939&\textbf{24.189}/0.935&\underline{24.069}/\textbf{0.943}&23.641/0.940\\
    \cline{2-10}
   & 71.201 & 54.381 & 7.806 &7.689 & 9.580 & 102.513 & 56.960 &33.768&17.689\\
   \cline{1-10}
    \multirow{2}*{Image(4)}&21.117/0.915&\underline{23.952}/0.954&23.946/0.956&\underline{23.950/0.957}&23.637/0.954&23.463/0.951&23.817/0.954&\textbf{24.197/0.959}&23.948/\underline{0.958}\\
    \cline{2-10}
   & 71.170 & 52.891 & 7.720 &7.889 & 9.668 & 83.029 & 53.263 &32.669&17.483\\
   \cline{1-10}
    \multirow{2}*{Image(5)}&28.176/0.931&\underline{30.510}/0.951&30.229/\underline{0.966}&30.391/0.966&29.623/0.962&29.001/0.941&29.201/0.948&\textbf{30.755/0.968}&29.949/0.961\\
     \cline{2-10}
   & 71.223 & 45.694 & 7.826 &7.819 & 9.431 & 87.132 & 51.844 &32.187&16.203\\
   \cline{1-10}
    \multirow{2}*{Image(6)}&23.599/0.960&24.769/0.969&25.240/0.975&25.326/0.975&24.526/0.969&25.596/0.975&\underline{25.833}/0.976&\textbf{26.265/0.980}&25.819/\underline{0.977}\\
    \cline{2-10}
   & 73.157 & 89.845 & 7.977 &8.144 & 11.964 & 94.439 & 53.857 &42.806&18.568\\
   \cline{1-10}
    \multirow{2}*{Image(7)}&25.177/0.963&26.577/0.975&26.856/0.978&\underline{27.083/0.979}&26.326/0.975&26.744/0.976&26.936/0.977&\textbf{27.644/0.980}&26.901/0.978\\
    \cline{2-10}
   & 71.846 & 40.277 & 7.821 &7.955 & 9.688 & 92.548 & 55.923 &32.161&19.583\\
   \cline{1-10}
   \multirow{2}*{Image(8)}&21.613/0.920&24.388/0.956&24.574/0.966&24.340/0.964&24.246/0.962&24.075/0.959&24.022/0.960&\textbf{24.641/0.969}&\underline{24.593/0.965}\\
   \cline{2-10}
   & 70.038 & 40.180 & 7.803 &7.854 & 9.552 & 94.098 & 53.030&21.664&17.020\\
   \cline{1-10}
   \multirow{2}*{Image(9)}&22.100/0.831&24.079/0.879&24.536/0.907&24.714/0.909&23.871/0.896&24.459/0.901&\underline{24.811}/0.902&24.600/\underline{0.910}&\textbf{24.828/0.911}\\
   \cline{2-10}
   & 74.518 & 42.610 & 7.650 &7.916 & 9.640 & 101.516 & 55.032&34.078&19.440\\
    \bottomrule
\label{E4}
    \end{tabular}}
    \end{threeparttable}
\end{table*}

\begin{figure*}
\begin{floatrow}[2]
\figurebox{\caption{The first column is the original images, the second column is the observed images, the third column to the last column is the recovering results by LRQA-2, Q-DNN, QTNN, and DWQTNN, orderly.}}{
\label{E0.65}%
  \includegraphics[width=6cm]{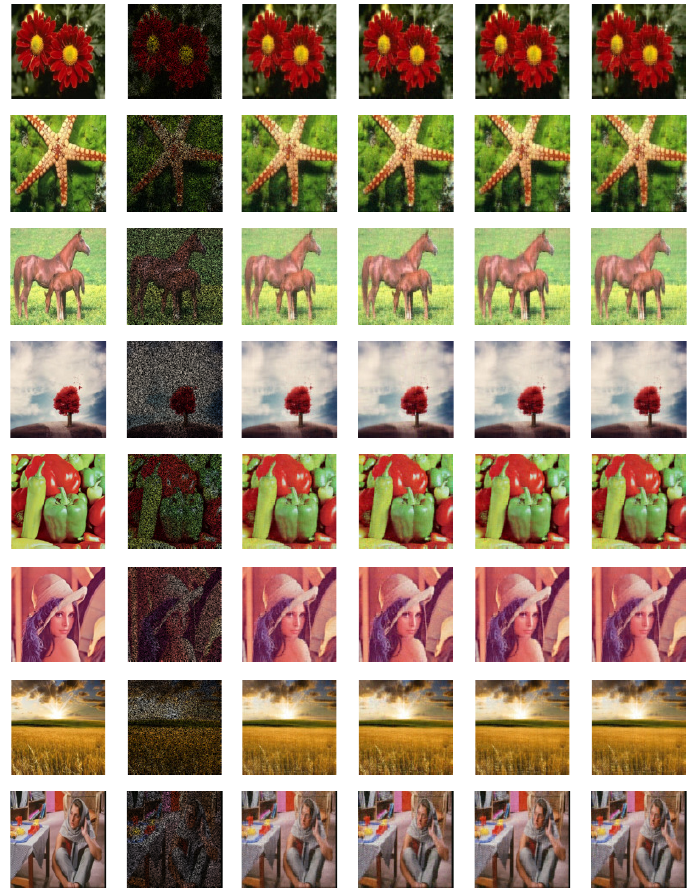}}%
\tablebox{\caption{QUANTITATIVE EVALUATION (PSNR/SSIM and TIME Consumption(s)) OF QUATERNION-BASED COLOR IMAGE COMPLETION ALGORITHMS}}{%
\scalebox{0.5}{
\renewcommand\arraystretch{1.75}
  \begin{tabular}{|c|c|c|c|c|}
  \cline{1-5}
    \multirow{2}*{Method}&
    \multicolumn{4}{c|}{\emph{p}=0.65}\\
    \cline{2-5}
    &LRQA-2 \cite{8844978}&Q-DNN \cite{9204671}&QTNN [ours]&DWQTNN [ours]\\
     \cline{1-5}
    \multirow{2}*{Image(2)}&25.188/0.924&25.230/0.924&\textbf{25.378/0.930}&24.648/0.918\\
    \cline{2-5}
   & 178.770 & 70.017 & 57.740 &19.050 \\ \hline
    \cline{1-5}
    \multirow{2}*{Image(3)}&21.739/0.887&21.862/0.883&\textbf{21.925/0.895}&21.562/0.890\\
    \cline{2-5}
   & 116.549 & 60.435 & 48.449 &18.179\\
   \cline{1-5}
   \multirow{2}*{Image(4)}&22.280/0.919&22.183/0.916&\textbf{22.683/0.927}&22.611/0.926\\
    \cline{2-5}
   & 104.655 & 65.918 & 39.339 &19.026 \\
   \cline{1-5}
    \multirow{2}*{Image(5)}&28.087/0.917&28.260/0.919&\textbf{28.767/0.938}&28.321/0.931\\
    \cline{2-5}
   & 162.551 & 74.765 & 56.066&26.881\\
   \cline{1-5}
    \multirow{2}*{Image(6)}&23.143/0.946&23.388/0.948&\textbf{23.400/0.951}&23.041/0.946\\
    \cline{2-5}
   & 150.949 & 80.802 & 46.156&20.689 \\
   \cline{1-5}
   \multirow{2}*{Image(7)}&24.667/0.954&\textbf{24.928}/0.955&24.904/\textbf{0.957}&24.344/0.954\\
    \cline{2-5}
   & 113.901 & 69.019& 37.439&16.470 \\
   \cline{1-5}
    \multirow{2}*{Image(8)}&23.120/0.932&22.983/0.931&\textbf{23.411/0.938}&23.369/0.937\\
    \cline{2-5}
   & 135.958 & 70.766& 38.237&16.300 \\
   \cline{1-5}
  \multirow{2}*{Image(9)}&22.578/0.826&22.656/0.825&\textbf{22.781/0.835}&22.742/0.833\\
    \cline{2-5}
   & 114.322 & 57.731 & 49.810 &17.577 \\
   \cline{1-5}
  \end{tabular} \label{E5}}}
\end{floatrow}
\end{figure*}

\begin{figure*}
\begin{floatrow}[2]
\figurebox{\caption{The first column is the original images, the second column is the observed images, the third column to the last column is the recovering results by LRQA-2, Q-DNN, QTNN, and DWQTNN, orderly.}}{
\label{E0.75}%
  \includegraphics[width=6cm]{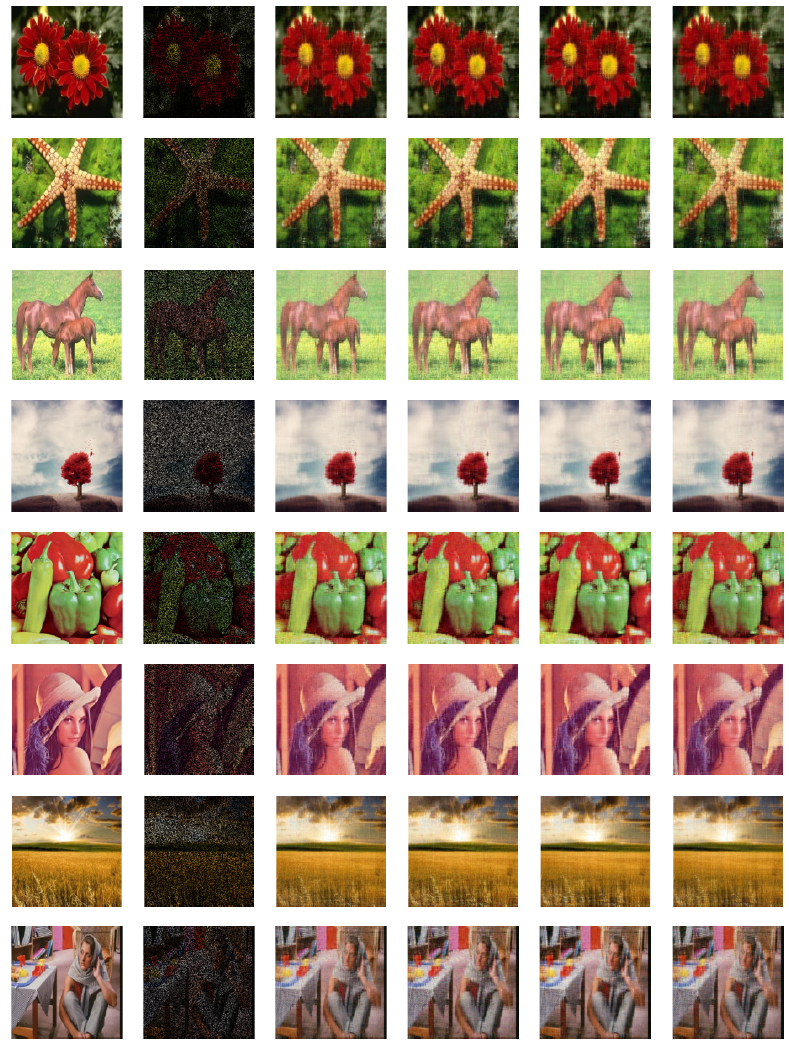}}
\tablebox{\caption{QUANTITATIVE EVALUATION (PSNR/SSIM and TIME Consumption(s)) OF QUATERNION-BASED COLOR IMAGE COMPLETION ALGORITHMS}}{%
\scalebox{0.5}{
\renewcommand\arraystretch{1.75}
  \begin{tabular}{|c|c|c|c|c|}
  \cline{1-5}
    \multirow{2}*{Method}&
    \multicolumn{4}{c|}{\emph{p}=0.75}\\
    \cline{2-5}
    &LRQA-2 \cite{8844978}&Q-DNN \cite{9204671}&QTNN [ours]&DWQTNN [ours]\\
    \cline{1-5}
    \multirow{2}*{Image(2)}&23.374/0.871&\textbf{23.522}/0.870&23.498/\textbf{0.872}&22.691/0.858\\
    \cline{2-5}
   & 153.130 & 59.075 & 61.628 &16.450 \\
   \cline{1-5}
    \multirow{2}*{Image(3)}&20.088/0.829&20.027/0.819&\textbf{20.072/0.834}&19.700/0.827\\
    \cline{2-5}
   & 134.148 & 57.896& 60.522 &17.867 \\
   \cline{1-5}
   \multirow{2}*{Image(4)}&21.243/0.887&21.123/0.881&\textbf{21.512/0.894}&21.336/0.891\\
    \cline{2-5}
   & 123.904 & 68.170 & 38.918 &18.134 \\
   \cline{1-5}
    \multirow{2}*{Image(5)}&27.038/0.882&26.953/0.881&\textbf{27.251/0.897}&26.631/0.881\\
    \cline{2-5}
   & 140.097 & 52.491 & 39.631 &16.545 \\
   \cline{1-5}
    \multirow{2}*{Image(6)}&21.249/0.911&\textbf{21.302}/0.910&21.241/\textbf{0.914}&20.200/0.891\\
    \cline{2-5}
   & 130.026 & 70.586 & 49.177&16.910 \\
   \cline{1-5}
   \multirow{2}*{Image(7)}&22.884/\textbf{0.927}&\textbf{23.046/0.927}&22.870/\textbf{0.927}&22.363/0.924\\
    \cline{2-5}
   & 131.112 & 72.024 & 51.988 &16.797 \\
   \cline{1-5}
    \multirow{2}*{Image(8)}&22.404/0.907&22.160/0.905&\textbf{22.503/0.912}&22.507/0.909\\
    \cline{2-5}
   & 143.342 & 66.836 & 39.396&16.211 \\
   \cline{1-5}
  \multirow{2}*{Image(9)}&20.981/0.745&20.906/0.737&\textbf{20.983/0.745}&20.271/0.717\\
    \cline{2-5}
   & 126.937 & 56.608 & 58.938 &16.665 \\
   \cline{1-5}
  \end{tabular}\label{E6}}}
\end{floatrow}
\end{figure*}

\section{Conclusion}
\label{Conclusion}
We propose two LRQMC algorithms in this paper that are based on the quaternion-based truncated nuclear norm (QTNN) to depict the low-rank property. Although these method base on quaternion nuclear norm (QNN), the truncated strategy makes the calculation of QSVD only need several times, such that the time consumption would be less. For QTNN, we utilize the two-step strategy, and the ADMM framework is used to solve this model in the second step. For DWQTNNR, we derive one-step optimization to make the quaternion truncated nuclear norm method be more efficient to deal with block missing images and faster when dealing with random missing images with acceptable indexes. Experimental results prove that our method can achieve a developed performance.

Because the most time-consuming process is to calculate QSVD, so one of our future work is to try to find a more efficient method to improve this calculation and also depict the low-rank property accurately e.g., taking advantage of deep learning. Besides, our weighted method could handle block missing images well, however when solving the random missing problems, it would be faster and the indexes would be underperformance. Hence, the other future work is to improve the performance of these methods from the point of quaternion computation and utilize more advantages of quaternion-based representation.

\appendix
\section{The PROOF OF Theorem \ref{theorem3}  }
\label{A1}
Proof: Assume that $M\leq N$, and the QSVD of the quaternion matrix $\dot{\textbf{X}}\in \mathbb{H}^{M\times N}$ can be represented as:
\begin{equation}
\dot{\mathbf{X}}={\dot{\mathbf{U}}
\left( \begin{array}{cc}
\mathbf{\Sigma}_{\tilde{r}} & \mathbf{0}  \\
\mathbf{0} & \mathbf{0}\\
\end{array}
\right )\dot{\mathbf{V}}^H}=\dot{\mathbf{U}}\mathbf{D}\dot{\mathbf{V}}^H
\end{equation}
where $\mathbf{\Sigma}_r=diag({\sigma_1,\cdots, \sigma_{\tilde{r}}})\in\mathbb{R}^{\tilde{r}\times \tilde{r}}$, and all singular values $\sigma_i (i=1,\cdots,r)$ are nonnegative, $\dot{\textbf{U}}\in\mathbb{H}^{M \times M}$ and $\dot{\mathbf{V}}\in\mathbb{H}^{N \times N}$ are two unitary quaternion matrices.
Then,
\begin{equation}
\mid tr(\dot{\mathbf{A}}\dot{\mathbf{X}}\dot{\mathbf{B}}^{H})\mid=
\mid tr(\dot{\mathbf{A}}\dot{\mathbf{U}}\mathbf{D}\dot{\mathbf{V}}^H\dot{\mathbf{B}}^{H})\mid
\end{equation}
Let $\dot{\mathbf{U}}_0=\dot{\mathbf{A}}\dot{\mathbf{U}}=(\dot{u}_{ij})\in\mathbb{H}^{r \times M}$ and $\dot{\mathbf{V}}_0=\dot{\mathbf{B}}\dot{\mathbf{V}}=(\dot{v}_{ij})\in\mathbb{H}^{r \times N}$, distinctly,
$\dot{\mathbf{U}}_0\dot{\mathbf{U}}_0^H=\mathbf{I}_{r\times r}$ and $\dot{\mathbf{V}}_0\dot{\mathbf{V}}_0^H=\mathbf{I}_{r\times r}$. Then we have
\begin{equation}
\begin{aligned}
&\mid tr(\dot{\mathbf{A}}\dot{\mathbf{X}}\dot{\mathbf{B}}^{H})\mid=\mid\dot{\mathbf{U}}_0\mathbf{D}\dot{\mathbf{V}}_0^H\mid
=\mid\sum_{i=1}^r\sum_{j=1}^M\sigma_j\dot{u}_{ij}\bar{\dot{v}}_{ij}\mid
\\&\leq\sum_{i=1}^r\sum_{j=1}^M\sigma_j\mid\dot{u}_{ij}\bar{\dot{v}}_{ij}\mid
=\mid(1,\cdots,1)_{1\times r}\mathbf{P}_{r\times M}(\sigma_i,\cdots,\sigma_M)^T\mid
\\&=\mid(1,\cdots,1,0,\cdots,0)_{1\times M}
\left(
\begin{array}{cc}
\mathbf{P}_{r\times M}  \\
\mathbf{0}_{(M-r)\times M} \\
\end{array}
\right)
 (\sigma_i,\cdots,\sigma_M)^T\mid
\end{aligned}
\end{equation}
where $\mathbf{P}_{r\times M}=(\mid\dot{u}_{ij}\bar{\dot{v}}_{ij}\mid)_{r\times M}$.
Because
\begin{equation}
\sum_{i=1}^r\mid\dot{u}_{ij}\bar{\dot{v}}_{ij}\mid\leq\frac{1}{2}[\sum_{i=1}^r\mid\dot{u}_{ij}\mid^2+\sum_{i=1}^r\mid\bar{\dot{v}}_{ij}\mid^2]=1
\end{equation}
\begin{equation}
\sum_{j=1}^M\mid\dot{u}_{ij}\bar{\dot{v}}_{ij}\mid\leq\frac{1}{2}[\sum_{j=1}^M\mid\dot{u}_{ij}\mid^2+\sum_{j=1}^M\mid\bar{\dot{v}}_{ij}\mid^2]\leq1
\end{equation}
, $\mathbf{P}_{r\times M} $ is doubly-substochastic matrix. According to the theory in \cite{marshall1979inequalities} (P 136 H.3.b) and \cite{horn2012matrix} (Theorem 8.1.4 and Theorem 8.7.6), we have
\begin{equation}
\begin{aligned}
&\mid tr(\dot{\mathbf{A}}\dot{\mathbf{X}}\dot{\mathbf{B}}^{H})\mid
\\&\leq\mid(1,\cdots,1,0,\cdots,0)_{1\times M}
\left(
\begin{array}{cc}
\mathbf{P}_{r\times M}  \\
\mathbf{0}_{(M-r)\times M} \\
\end{array}
\right)
(\sigma_i,\cdots,\sigma_M)^T\mid
\\&\leq\mid(1,\cdots,1,0,\cdots,0)_{1\times M}(\sigma_i,\cdots,\sigma_M)^T\mid
\\&=\sum_{i=1}^r\sigma_i.
\end{aligned}
\end{equation}
When $\dot{\mathbf{A}}= [\mathbf{I}_{r\times r},\mathbf{0}_{(r)\times (M-r)} ]\dot{\mathbf{U}}^H$, and
$\dot{\mathbf{B}}= [\mathbf{I}_{r\times r},\mathbf{0}_{(r)\times (N-r)} ]\dot{\mathbf{V}}^H$, we get
\begin{equation}
\max |tr(\dot{\mathbf{A}}\dot{\mathbf{X}}\dot{\mathbf{B}}^H)|=\sum_{i=1}^r\sigma_{i}(\dot{\mathbf{X}}).
\end{equation}
\section{The PROOF OF Theorem \ref{theorem4}}
\label{A2}
The proof of Theorem \ref{theorem4} is similar with that in the real matrix domain \cite{liu2015truncated}. We demonstrate that the mathematical details also hold in quaternion domain as follows.

When the step size of Eq. (\ref{model30}) changed to a smaller positive number  $\gamma_k$ $(0<\gamma_k<\beta_k)$, besides, $\gamma_k$ satisfies: $\sum_{k=1}^{+\infty}\frac{\gamma_k}{\beta_k}=c$ (c is a positive constant). Eqs. (\ref{model27}), (\ref{model29}-\ref{model31}) are reformulated as:
 \begin{equation}\label{modelB1}
\dot{\mathbf{H}}_{k+1}=\dot{\mathbf{\mathbb{X}}}_{k}+\frac{1}{\beta_k}(\mathbf{W}^{-2}\dot{\mathbf{C}}_l^H\dot{\mathbf{D}}_l+\mathbf{W}^{-1}\dot{\mathbf{Y}}_k).
\end{equation}
 \begin{equation}\label{modelB2}
\dot{\mathbf{H}}_{k+1}=P_{\Omega^C}(\dot{\mathbf{H}}_{k+1})+P_\Omega(\dot{\mathbf{M}}).
\end{equation}

\begin{equation}\label{modelB3}
\dot{\mathbf{\mathbb{X}}}_{k+1}=
\dot{\mathbf{H}}_{k+1}-\frac{1}{\beta_k}(\mathbf{W}^{-2}\dot{\mathbf{A}}_l^H\dot{\mathbf{B}}_l+\mathbf{W}^{-1}\dot{\mathbf{Y}}_k).
\end{equation}

\begin{equation}\label{modelB4}
\dot{\mathbf{Y}}_{k+1}=\dot{\mathbf{Y}}_{k}+\gamma_k\mathbf{W}(\dot{\mathbf{\mathbb{X}}}_{k+1}-\dot{\mathbf{H}}_{k+1}),
\end{equation}
Basing on the update of  $\dot{\mathbf{Y}}_{k}$ (\ref{modelB4}) and (\ref{modelB1}), (\ref{modelB3}), for the consequence of ${\dot{\mathbf{Y}}_{k}}$, we have
\begin{equation}\label{m1}
\begin{aligned}
\dot{\mathbf{Y}}_{k+1}&=\dot{\mathbf{Y}}_{k}+\gamma_k\mathbf{W}(\dot{\mathbf{\mathbb{X}}}_{k+1}-\dot{\mathbf{H}}_{k+1})
\\&=\dot{\mathbf{Y}}_{k}-\frac{\gamma_k}{\beta_k}(\mathbf{W}^{-1}\dot{\mathbf{A}}_l^H\dot{\mathbf{B}}_l+\dot{\mathbf{Y}}_{k})
\\&=(1-\frac{\gamma_k}{\beta_k})\dot{\mathbf{Y}}_{k}-\frac{\gamma_k}{\beta_k}\mathbf{W}^{-1}\dot{\mathbf{A}}_l^H\dot{\mathbf{B}}_l.
\end{aligned}
\end{equation}
Let $s_k=\frac{\gamma_k}{\beta_k}$, $d_k=1-s_k$, and $\dot{\mathbf{T}}=\mathbf{W}^{-1}\dot{\mathbf{A}}_l^H\dot{\mathbf{B}}_l$. (\ref{m1}) can be rewritten as:
\begin{equation}\label{m2}
\begin{aligned}
\dot{\mathbf{Y}}_{k+1}&=d_k\dot{\mathbf{Y}}_{k}-s_k\dot{\mathbf{T}}
\\&=d_k(d_{k-1}\dot{\mathbf{Y}}_{k-1}-s_{k-1}\dot{\mathbf{T}})-s_k\dot{\mathbf{T}}
\\&=\prod_{i=1}^kd_i\dot{\mathbf{Y}}_1-(\sum_{i=1}^{k-1}(s_i\prod_{j=i+1}^kd_j)\dot{\mathbf{T}}-s_k\dot{\mathbf{T}}.
\end{aligned}
\end{equation}
Because $0<s_k<1$, then we have
\begin{equation}\label{m3}
\begin{aligned}
\parallel\dot{\mathbf{Y}}_{k+1}\parallel_F&=\parallel\prod_{i=1}^kd_i\dot{\mathbf{Y}}_1-\sum_{i=1}^{k-1}(s_i\prod_{j=i+1}^kd_j)\dot{\mathbf{T}}\parallel_F
\\&\leq\parallel\dot{\mathbf{Y}}_1\parallel_F+\sum_{i=1}^{k}s_i\parallel\dot{\mathbf{T}}\parallel_F.
\end{aligned}
\end{equation}
Besides, $\sum_{k=1}^{+\infty}\frac{\gamma_k}{\beta_k}=c$ , it means that $\lim_{k\rightarrow\infty}s_k=0$ and  the sequence ${\parallel\dot{\mathbf{Y}}_{k}\parallel_F}$ produced by (\ref{modelB1}-\ref{modelB4}) has a upper bound.
\begin{equation}\label{m4}
\begin{aligned}
\parallel\dot{\mathbf{Y}}_{k}\parallel_F&\leq(\parallel\dot{\mathbf{Y}}_1\parallel_F+c\parallel\dot{\mathbf{T}}\parallel_F)=\parallel\dot{\mathbf{Y}}\parallel_F^{sup}.
\end{aligned}
\end{equation}
Hence, we can rewrite (\ref{m1}) as follows:
\begin{equation}\label{m5}
\dot{\mathbf{Y}}_{k+1}-\dot{\mathbf{Y}}_{k}=-s_k\dot{\mathbf{Y}}_{k}-s_k\dot{\mathbf{T}}
\end{equation}
\begin{equation}\label{m6}
\parallel\dot{\mathbf{Y}}_{k+1}-\dot{\mathbf{Y}}_{k}\parallel_F\leq s_k(\parallel\dot{\mathbf{Y}}\parallel_F^{sup}+\parallel\dot{\mathbf{T}}\parallel_F)
\end{equation}
\begin{equation}\label{m7}
\lim_{k\rightarrow\infty}\parallel\dot{\mathbf{Y}}_{k+1}-\dot{\mathbf{Y}}_{k}\parallel_F=0
\end{equation}
Basing on the update of  $\dot{\mathbf{H}}_{k}$ (\ref{modelB3}) and (\ref{modelB1}), for the consequence of ${\dot{\mathbf{H}}_{k}}$, we have
\begin{equation}\label{m8}
\begin{aligned}
\dot{\mathbf{H}}_{k+1}-\dot{\mathbf{H}}_{k}=&
-\frac{1}{\beta_{k-1}}(\mathbf{W}^{-2}\dot{\mathbf{A}}_l^H\dot{\mathbf{B}}_l+\mathbf{W}^{-1}\dot{\mathbf{Y}}_{k-1})
\\&+\frac{1}{\beta_k}(\mathbf{W}^{-2}\dot{\mathbf{C}}_l^H\dot{\mathbf{D}}_l+\mathbf{W}^{-1}\dot{\mathbf{Y}}_k).
\end{aligned}
\end{equation}
Because the step size $\beta_k>\beta_{k-1}>0$, if $\lim_{k\rightarrow\infty}\frac{1}{\beta_k}=0$, we have
\begin{equation}\label{m9}
\begin{aligned}
\|\dot{\mathbf{H}}_{k+1}-\dot{\mathbf{H}}_{k}\|_F\leq&
\frac{1}{\beta_{k-1}}\parallel\mathbf{W}^{-2}(\dot{\mathbf{C}}_l^H\dot{\mathbf{D}}_l-\dot{\mathbf{A}}_l^H\dot{\mathbf{B}}_l)\parallel_F
\\&+\frac{1}{\beta_k}\parallel\mathbf{W}^{-1}(\dot{\mathbf{Y}}_k-\dot{\mathbf{Y}}_{k-1})\parallel_F.
\end{aligned}
\end{equation}
\begin{equation}\label{m10}
\lim_{k\rightarrow\infty}\|\dot{\mathbf{H}}_{k+1}-\dot{\mathbf{H}}_{k}\|_F=0
\end{equation}
Because the Frobenius norm also belongs to the unitarily invariant norm in the quaternion domain \cite{jiang2003equality}. So $\|\dot{\mathbf{A}}_l^H\dot{\mathbf{B}}_l\|_F=\sqrt{M}$, and $\| \dot{\mathbf{C}}_l^H\dot{\mathbf{D}}_l\|_F=\sqrt{r}$, for $l=1,2,\cdots,$ and $r$ is the truncated number.

Then, by substituting $\dot{\mathbf{H}}_{k+1}$ into $\dot{\mathbf{\mathbb{X}}}_{k+1}$, we get:
\begin{equation}\label{modelB5}
\dot{\mathbf{\mathbb{X}}}_{k+1}=
\dot{\mathbf{\mathbb{X}}}_k-\frac{1}{\beta_k}(\mathbf{W}^{-2}\dot{\mathbf{A}}_l^H\dot{\mathbf{B}}_l+\mathbf{W}^{-2}\dot{\mathbf{C}}_l^H\dot{\mathbf{D}}_l).
\end{equation}
\begin{equation}\label{modelB6}
\begin{aligned}
\|\dot{\mathbf{\mathbb{X}}}_{k+1}-\dot{\mathbf{\mathbb{X}}}_k\|_F&=
\frac{1}{\beta_k}\|(\mathbf{W}^{-2}\dot{\mathbf{A}}_l^H\dot{\mathbf{B}}_l+\mathbf{W}^{-2}\dot{\mathbf{C}}_l^H\dot{\mathbf{D}}_l\|_F)
\\&\leq\frac{1}{\beta_k}\|\mathbf{W}^{-2}\|_F(\|\dot{\mathbf{A}}_l^H\dot{\mathbf{B}}_l\|_F+\|\dot{\mathbf{C}}_l^H\dot{\mathbf{D}}_l\|_F)
\\&\leq\frac{1}{\beta_k}\|\mathbf{W}^{-2}\|_F(\sqrt{M}+\sqrt{r}),
\end{aligned}
\end{equation}
where r is the truncated number of the singular values.
If $\lim_{k\rightarrow\infty}\frac{1}{\beta_k}=0$, i.e. $\beta_k$ increase progressively, then we have:
\begin{equation}\label{modelB7}
\lim_{k\rightarrow\infty}\|\dot{\mathbf{\mathbb{X}}}_{k+1}-\dot{\mathbf{\mathbb{X}}}_k\|_F=0.
\end{equation}
Hence, the sequence ${\dot{\mathbf{\mathbb{X}}}_k,  k =1,2,\cdots}$ derived from (\ref{modelB1}) and (\ref{modelB3}) will convergence.

Besides, the Lagrange function is convex function for $\dot{\mathbf{\mathbb{X}}}$ and $\dot{\mathbf{H}}$ with a fixed $\dot{\mathbf{Y}}$, the sequence ${\dot{\mathbf{\mathbb{X}}}_k,  k =1,2,\cdots}$ will converge to its local optimal solution, which is same with that in the real domain. If it consumes N iterations for the converge. From (\ref{modelB5}), we can get
\begin{equation}\label{modelB8}
\dot{\mathbf{X}}_{N}=
\dot{\mathbf{X}}_l+\sum_{i=1}^{N-1}\frac{1}{\beta_k}(\mathbf{W}^{-2}\dot{\mathbf{C}}_l^H\dot{\mathbf{D}}_l-\mathbf{W}^{-2}\dot{\mathbf{A}}_l^H\dot{\mathbf{B}}_l),
\end{equation}
where $\dot{\mathbf{X}}_l=\dot{\mathbf{\mathbb{X}}}_1$.
Then, combining (\ref{modelB1}) and (\ref{modelB4}), we have
\begin{equation}\label{modelB9}
\begin{aligned}
&\|P_\Omega(\dot{\mathbf{\mathbb{X}}}_{k+1})-P_\Omega(\dot{\mathbf{M}})\|_F=\\&\|
-\frac{1}{\beta_k}P_\Omega(\mathbf{W}^{-2}\dot{\mathbf{A}}_l^H\dot{\mathbf{B}}_l+\mathbf{W}^{-1}\dot{\mathbf{H}}_{k})\|_F.
\end{aligned}
\end{equation}
Because $\lim_{k\rightarrow\infty}\frac{1}{\beta_k}=0$, then the following equation will hold
\begin{equation}\label{modelB91}
\lim_{k\rightarrow\infty}\|P_\Omega(\dot{\mathbf{\mathbb{X}}}_{k+1})-P_\Omega(\dot{\mathbf{M}})\|_F=0
\end{equation}
\begin{equation}\label{modelB92}
\|P_\Omega(\dot{\mathbf{X}}_{N})-P_\Omega(\dot{\mathbf{M}})\|_F=0
\end{equation}
Let $\frac{1}{\epsilon_l}=\sum_{i=1}^{N-1}\frac{1}{\beta_k}$, then the update of $\dot{\mathbf{X}}_{N}$ can be reformulated as:
\begin{equation}\label{modelB13}
\dot{\mathbf{X}}_{N}=
\dot{\mathbf{X}}_l+\frac{1}{\epsilon_l}(\mathbf{W}^{-2}\dot{\mathbf{C}}_l^H\dot{\mathbf{D}}_l-\mathbf{W}^{-2}\dot{\mathbf{A}}_l^H\dot{\mathbf{B}}_l),
\end{equation}
where $\dot{\mathbf{X}}_l=\dot{\mathbf{\mathbb{X}}}_1$.
\begin{equation}\label{modelB14}
\dot{\mathbf{X}}_{N}= P_\Omega^C(\dot{\mathbf{X_{N}}})+P_\Omega(\dot{\mathbf{M}}).
\end{equation}
where $\dot{\mathbf{X}}_l=\dot{\mathbf{\mathbb{X}}}_1$.
\section{The PROOF OF Theorem \ref{theorem5}}
\label{A3}
Eq. (\ref{model41}) can be represented as:
\begin{equation}\label{modelB15}
\dot{\mathbf{X}}_{k+1}-\dot{\mathbf{X}}_{k}=
\frac{1}{\varepsilon_k}(\mathbf{W}_1\dot{\mathbf{A}}_k^H\dot{\mathbf{B}}_k
- \mathbf{W}_2\dot{\mathbf{C}}_k^H\dot{\mathbf{D}}_k).
\end{equation}
Then we have
\begin{equation}\label{modelB16}
\begin{aligned}
\|\dot{\mathbf{X}}_{k+1}-\dot{\mathbf{X}}_k\|_F&=
\frac{1}{\varepsilon_k}\|\mathbf{W}_1\dot{\mathbf{A}}_k^H\dot{\mathbf{B}}_k
- \mathbf{W}_2\dot{\mathbf{C}}_k^H\dot{\mathbf{D}}\|_k
\\&\leq\frac{1}{\varepsilon_k}(\|\mathbf{W}_1\|_F(\|\dot{\mathbf{A}}_k^H\dot{\mathbf{B}}_k\|_F)\\&\quad+
                               \|\mathbf{W}_2\|_F(\|\dot{\mathbf{C}}_k^H\dot{\mathbf{D}}_k\|_F))
\\&\leq\frac{1}{\varepsilon_k}(\|\mathbf{W}_1\|_F\sqrt{M}+\|\mathbf{W}_2\|_F\sqrt{r}).
\end{aligned}
\end{equation}
where  r is the truncated number of the singular values. Supposing that $\epsilon$ as one stop tolerance for the DWQTNN method. We have
\begin{equation}\label{modelB17}
\frac{1}{\varepsilon_k}(\|\mathbf{W}_1\|_F\sqrt{M}+\|\mathbf{W}_2\|_F\sqrt{r})\leq \epsilon.
\end{equation}
Besides, we define $\varepsilon_{k+1}=\rho\varepsilon_k$, where $\varepsilon_1>0, \rho>1$, and $k=1, 2, \cdots.$ in (\ref{model38}). So we have $\varepsilon_{k}=\rho^{k-1}\varepsilon_1$. Let $c=\|\mathbf{W}_1\|_F\sqrt{M}+\|\mathbf{W}_2\|_F\sqrt{r}$, then the inequality (\ref{modelB17}) can be rewritten as
\begin{equation}\label{modelB18}
\frac{1}{\rho^{k-1}\varepsilon_1}\ c \leq \epsilon.
\end{equation}
\begin{equation}\label{modelB19}
k\geq1-\frac{\ln(\varepsilon_1\epsilon)-\ln(c)}{\ln(\rho)}.
\end{equation}

\section*{Acknowledgements}
This work was supported by the Science and Technology Development Fund, Macau SAR (File no. FDCT/085/2018/A2) and University of Macau (File no. MYRG2019-00039-FST).





%
%
%
\bibliographystyle{unsrt}
\bibliography{mybibfile}
\end{document}